\documentclass[12pt]{article}

\usepackage{amsmath,amssymb}
\usepackage{hyperref}
\usepackage{verbatim}
\usepackage{stmaryrd}
\usepackage{amsfonts}

\usepackage{tikz}
\usetikzlibrary{matrix,arrows,shapes}
\newcommand{\btp}{\begin{tikzpicture}[baseline=-5pt,scale=0.25,line width=0.7pt]}
\newcommand{\etp}{\end{tikzpicture}}

\newcommand{\twobv}{\begin{array}{c}\boxslash\vspace{-.22cm}\end{array}\hspace{-0.68cm}\begin{array}{c}\boxslash\vspace{.25cm}\end{array}\!}
\newcommand{\twobh}{\boxslash\hspace{-.18cm}\boxslash}

\begin{document}


\begin{titlepage}

\begin{centering}
\vspace{.2in}

{\Large {\bf Integrable achiral $D5$-brane reflections and asymptotic Bethe equations}}

\vspace{.3in}

{\large Diego H. Correa${}^{a,1}$, Vidas Regelskis${}^{2,b,c}$ and C. A. S. Young${}^{b,3}$}\\
\vspace{.2 in}
${}^{a}${\emph{IFLP-CONICET and Departamento de F\'{\i}sica,\\
Facultad de Ciencias Exactas, Universidad Nacional de La Plata,\\
C.C. 67, (1900) La Plata, Argentina}} \\
${}^{b}${\emph{Department of Mathematics, University of York,\\
Heslington, York YO10 5DD, UK}}\\
${}^{c}${\emph{Institute of Theoretical Physics and Astronomy of Vilnius University,\\
Go\v{s}tauto 12, Vilnius 01108, Lithuania}}\\
%

\footnotetext[1]{{\tt correa@fisica.unlp.edu.ar,}\quad ${}^{2}${\tt vr509@york.ac.uk}\quad ${}^{2}${\tt charles.young@york.ac.uk}}
\vspace{.5in}

{\bf Abstract}

\vspace{.1in}
\end{centering}

{We study the reflection of magnons from a $D5$-brane in the framework of the AdS/CFT correspondence. We consider two possible orientations of the $D5$-brane with respect to the reference vacuum state, namely vacuum states aligned along ``vertical'' and ``horizontal'' directions. We show that the reflections are of the achiral type. We also show that the reflection matrices satisfy the boundary Yang-Baxter equations for both orientations. In the horizontal case the reflection matrix can be interpreted in terms of a bulk S-matrix, $S(p,-p)$, and factorizability of boundary scattering therefore follows from that of bulk scattering. 
Finally, we solve the nested coordinate Bethe ansatz for the system in the vertical case to find the Bethe equations. In the horizontal case, the Bethe equations are of the same form as those for the closed string.}

\end{titlepage}

\tableofcontents


\section{Introduction}
Integrable QFTs with boundaries have long been studied because of their applications in many problems in physics. Over the past few years, certain nonlinear $\sigma$-models corresponding to strings propagating in AdS spacetimes have been the subject of intense investigations because of their importance for the understanding of the AdS/CFT correspondence. The AdS/CFT correspondence is a duality between certain seemingly different theories \cite{adscftmalda}. It states, for example, that the spectrum of scale dimensions in the conformal ${\cal N}=4$ super Yang-Mills theory in 4 dimensions coincides with the spectrum of energies of strings propagating in $AdS_5\times S^5$. For the full range of values of the interacting CFT coupling constant, this spectral problem is believed to be exactly integrable in the planar limit, i.e. in the limit of the gauge group rank $N\to \infty$. With the use of integrability methods substantial and steady progress has been observed in the resolution of this planar spectral problem (see \cite{overview} for a complete overview). Though most studies deal with periodic boundary conditions, cases with open boundary conditions have also been considered (see \cite{Zoubos:2010kh} and references therein). In some cases, the addition of open boundaries enriches the problem, for example, by making it less supersymmetric and by adding matter in the fundamental representation of the gauge group.

Supersymmetric  $D$3, $D$5 and $D$7-branes are natural candidates to introduce Dirichlet boundary conditions in the nonlinear
$\sigma$-model for strings on $AdS_5\times S^5$. An all-loop derivation of the reflection matrix for certain
$D$3-branes was performed in \cite{HMopen}, where the authors also showed that the boundary Yang-Baxter equation (bYBE) is satisfied. This was  in agreement with explicit weak and strong coupling limits where integrability had been previously observed \cite{BeVa,MaVa}. $D$7-branes also provide integrable boundary conditions \cite{Correa1}.

However, the integrability of the supersymmetric $D$5-brane boundary was less clear. On the one hand, 1-loop results on the gauge theory side \cite{DeMa} indicated that the dilatation operator in the scalar sector was an integrable Hamiltonian. On the other hand, the same procedure successfully used to show the classical integrability of the $D$3 and $D$7-brane boundary conditions \cite{MaVa}, could not be used to construct the infinite set of non-local charges for the case of the $D$5-brane\footnote{For the $D$5-brane, the charges could be constructed in an $SU(2)$ sub-sector only.}. This made it seem questionable whether integrability was present beyond 1-loop, though it by no means ruled it out. In part, it was this uncertainty that motivated \cite{Correa1}, where the all-loop reflection matrices for two possible orientations of the $D$5-brane were obtained. Unfortunately, that paper contained a sign error -- originating, as we discuss below, in a graded permutation -- which led to the erroneous conclusion that the bYBE was not fulfilled with $D$5-brane reflection matrices. In the present paper we revise that statement, to find not only that the $D$5-brane boundary conditions are integrable but also that they have additional interesting features. For instance, for one of the two possible $D$5-brane orientations, the reflection can be fully understood in terms of a bulk S-matrix of the form ${\cal S}(p,-p)$. Another interesting aspect is the achiral nature of the boundary reflection i.e. the incoming left particles become right ones after the reflection and the right particles become left ones \cite{MS03}. These new features play an important role in constructing a nested Bethe ansatz leading to the Bethe equations to solve the spectral problem with $D$5-brane boundary conditions.

The structure of this paper is as follows. In section \ref{Ref matrices} we review the all-loop $D$5-brane reflection matrices for the two possible relative orientations between the brane and the polarization of the vacuum. For the horizontal vacuum orientation we use a basis for the vector representation different than the one used in \cite{Correa1}, for which it becomes evident that the reflection matrices for a right boundary can be understood as a bulk S-matrices of the form ${\cal S}(p,-p)$. The factorizability of the bulk S-matrix constitutes a strong hint for the factorizability of the reflection matrix. We show that bYBE is fulfilled for both relative orientations and indicate what the error was in \cite{Correa1}. In section \ref{CBA} we formulate a nested Bethe ansatz for the vertical vacuum orientation, and derive the corresponding Bethe equations. We conclude in section \ref{discussion} with a discussion about our results and possible future directions.

\section{Reflection from the $D$5-brane}
\label{Ref matrices}
We begin by briefly recalling the setup and symmetries of the $D5$-brane as well as the
representations of the matter content living on the brane. The details can be found in \cite{Correa1}. We then present the corresponding reflection matrices, and show that they obey the boundary Yang-Baxter equation.

\subsection{Symmetries}
\noindent The symmetry algebra in the bulk of the scattering theory is $\mathfrak{psu}(2|2)\times\widetilde{\mathfrak{psu}}(2|2)\ltimes\mathbb{R}^{3}$,
consisting of two copies (left and right) of the centrally-extended algebra
$\mathfrak{psu}(2|2)\ltimes\mathbb{R}^3$ with their central charges identified. The generators of $\mathfrak{psu}(2|2)\ltimes\mathbb{R}^3$ are the central charges $\mathbb{H}$,
$\mathbb{C}$ and $\mathbb{C}^{\dagger}$, two sets of bosonic rotation
generators $\mathbb{R}_{a}^{\enskip b}$, $\mathbb{L}_{\alpha}^{\enskip\beta}$ and
two sets of fermionic supersymmetry generators $\mathbb{Q}_{\alpha}^{\enskip a},$
$\mathbb{G}_{a}^{\enskip\alpha}$. The non-trivial commutation
relations are \cite{Beisert1}
\begin{align}
 & \left[\mathbb{L}_{\enskip\alpha}^{\beta},\mathbb{J}^{\gamma}\right]=\delta_{\alpha}^{\gamma}\,\mathbb{J}^{\beta}-\frac{1}{2}\delta_{\alpha}^{\beta}\,\mathbb{J}^{\gamma}, &  & \left[\mathbb{L}_{\alpha}^{\enskip\beta},\mathbb{J}_{\gamma}\right]=-\delta_{\gamma}^{\beta}\,\mathbb{J}_{\alpha}+\frac{1}{2}\delta_{\alpha}^{\beta}\,\mathbb{J}_{\gamma},\nonumber \\
 & \left[\mathbb{R}_{\enskip a}^{b},\mathbb{J}^{c}\right]=\delta_{a}^{c}\,\mathbb{J}^{b}-\frac{1}{2}\delta_{a}^{b}\,\mathbb{J}^{c}, &  & \left[\mathbb{R}_{\enskip a}^{b},\mathbb{J}_{c}\right]=-\delta_{c}^{b}\,\mathbb{J}_{a}+\frac{1}{2}\delta_{a}^{b}\,\mathbb{J}_{c},\nonumber \\
 & \left\{ \mathbb{Q}_{\enskip a}^{\alpha},\mathbb{Q}_{\enskip b}^{\beta}\right\} =\epsilon_{ab}\epsilon^{\alpha\beta}\,\mathbb{C}, &  & \left\{ \mathbb{G}_{\enskip\alpha}^{a},\mathbb{G}_{\enskip\beta}^{b}\right\} =\epsilon_{\alpha\beta}\epsilon^{ab}\,\mathbb{C}^{\dagger},\nonumber \\
 & \left\{ \mathbb{Q}_{\enskip a}^{\alpha},\mathbb{G}_{\enskip\beta}^{b}\right\} =\delta_{a}^{b}\,\mathbb{L}_{\beta}^{\enskip\alpha}+\delta_{\beta}^{\alpha}\,\mathbb{R}_{a}^{\enskip b}+\frac{1}{2}\delta_{a}^{b}\delta_{\beta}^{\alpha}\,\mathbb{H},
\label{su22c}\end{align}
where $a,\; b,...=1,\;2$ and $\alpha,\;\beta,...=3,\;4$. We use undotted ($a$, $\alpha$) and dotted ($\dot{a}$, $\dot{\alpha}$) indices to distinguish generators of left and right $\mathfrak{psu}(2|2)$.

The kind of $D5$-brane we consider wraps an $AdS_{4}\subset AdS_{5}$
and a maximal $S^{2}\subset S^{5}$. The $AdS_4$ part of
the brane defines a $2+1$ dimensional defect hypersurface of the $3+1$
dimensional conformal boundary. The fundamental matter living on
the defect hypersurface is a 3d hypermultiplet \cite{DFO}.
The original $\mathfrak{so}(6)$ R-symmetry of $\mathcal N=4$ SYM is broken by the presence
of the $D$5-brane down to $\mathfrak{so}(3)_{H}\times \mathfrak{so}(3)_{V}$.
We shall fix the bulk vacuum state to be $Z=X^5+iX^6$ and consider two inequivalent embeddings of the $D5$-brane into $AdS_5 \times S^5$ in which the maximal $S^2\subset S^5$ is specified by:
\begin{itemize}
\item $X^{4}=X^{5}=X^{6}=0$, for which the vacuum is ``vertical'';
\item $X^{1}=X^{2}=X^{3}=0$, for which the vacuum is ``horizontal''.
\end{itemize}
At the boundary of the scattering theory, only those bulk symmetries that are also symmetries of the $D5$-brane are preserved.
The preserved symmetry algebra is a ``diagonal'' copy
\begin{equation}\nonumber \mathfrak{psu}(2|2)_D\ltimes\mathbb{R}^{3}\subset \mathfrak{psu}(2|2)\times\widetilde{\mathfrak{psu}}(2|2)\ltimes\mathbb{R}^{3},\end{equation} whose generators, obeying the canonical commutation relations (\ref{su22c}), we shall write as  $\check{\mathbb{L}}_{\;\check{\beta}}^{\check{\alpha}}$, $ \check{\mathbb{R}}_{\;\check{b}}^{\check{a}}$, $\check{\mathbb{Q}}_{\;\check{a}}^{\check{\alpha}}$, $\check{\mathbb{G}}_{\;\check{\alpha}}^{\check{a}}$ and $\check{\mathbb{H}}$,\;$\check{\mathbb{C}}$ and $\check{\mathbb{C}}^{\dagger}$. These generators are given by  $\check{\mathbb H} = \mathbb H + \tilde{\mathbb H}$,
$\check{\mathbb C} = \mathbb C + \kappa^2 \tilde{\mathbb C}$,
$\check{\mathbb C}^\dagger = \mathbb C^\dagger + \kappa^{-2} \tilde{\mathbb C}^\dagger$ and
\begin{align}
\check{\mathbb{L}}_{\;\check{\beta}}^{\check{\alpha}}
&=\mathbb{L}_{\;\beta}^{\alpha}+\tilde{\mathbb{L}}_{\;\bar{\dot{\beta}}}^{\bar{\dot{\alpha}}},
&\qquad
\check{\mathbb{R}}_{\;\check{b}}^{\check{a}}
&=\mathbb{R}_{\; b}^{a}+\tilde{\mathbb{R}}_{\;\dot{b}}^{\dot{a}};\label{LsD}\\
\check{\mathbb{Q}}_{\;\check{a}}^{\check{\alpha}}
&=\mathbb{Q}_{\; a}^{\alpha}+\kappa\,\tilde{\mathbb{Q}}_{\;\dot{a}}^{\bar{\dot{\alpha}}},
&\qquad\check{\mathbb{G}}_{\;\check{\alpha}}^{\check{a}}
&=\mathbb{G}_{\;\alpha}^{a}+\kappa^{-1}\,\tilde{\mathbb{G}}_{\;\bar{\dot{\alpha}}}^{\dot{a}},
\label{QsD}
\end{align}
where the bar above the dotted indices acts
as $\bar{\dot{3}}=\dot{4}$ and $\bar{\dot{4}}=\dot{3}$.
Here the number $\kappa$ depends on the orientation of the $D$5-brane \cite{Correa1}:
\begin{equation}
\kappa =
\begin{cases}
-i &\text{vertical case} \\
-1 & \text{horizontal case}.
\end{cases}
\end{equation}
The preserved $R$-symmetries $\check{\mathbb{R}}_{\;\check{b}}^{\check{a}}$ are the generators of $\mathfrak{so}(3)_H$ in the vertical case, and of $\mathfrak{so}(3)_V$ in the horizontal case.

\subsection{Bulk representation}
We need to determine how the elementary bulk magnons transform with respect to the preserved boundary symmetry algebra $\mathfrak{psu}(2|2)_{D}\ltimes\mathbb{R}^{3}$ generated by (\ref{LsD})-(\ref{QsD}). Recall that with respect to the bulk symmetry algebra $\mathfrak{psu}(2|2)\times\widetilde{\mathfrak{psu}}(2|2)\ltimes\mathbb{R}^{3}$, the bulk magnon transforms \cite{Beisert1} in the bifundamental representation  $(\boxslash_{(a,b,c,d)},\widetilde{\boxslash}_{(a,b,c,d)})$.
The representation labels $(a,b,c,d)$ are the same for both left and right factors, and are  conveniently parametrized by\footnote{Note that here we use parametrization different from the one used in \cite{Correa1}.}
\begin{equation}
a=\sqrt{\frac{g}{2}}\eta,\quad b=\sqrt{\frac{g}{2}}\frac{i\zeta}{\eta}\left(\frac{x^{+}}{x^{-}}-1\right),\quad c=-\sqrt{\frac{g}{2}}\frac{\eta}{\zeta x^{+}},\quad d=-\sqrt{\frac{g}{2}}\frac{x^{+}}{i\eta}\left(\frac{x^{-}}{x^{+}}-1\right),\label{abcd}
\end{equation}
where $\zeta= e^{2i\xi}$ is the magnon phase and unitarity requires $\eta= e^{i\xi} e^{i\frac{\varphi}{2}}\sqrt{i\left(x^{-}-x^{+}\right)}$.
The spectral parameters $x^\pm$ are constrained to satisfy the mass-shell condition
\begin{equation}
x^{+}+\frac{1}{x^{+}}-x^{-}-\frac{1}{x^{-}}=\frac{2 i}{g}.\label{mass-shell_1}
\end{equation}
The magnon momentum is $p$ where $e^{ip}=x^+/x^-$. We shall sometimes use the alternative notation $\mathcal{V}\left(p,\zeta\right)$ for the fundamental representation $\boxslash_{(a,b,c,d)}$.

It follows that, under the action of the diagonal subalgebra
$\mathfrak{psu}(2|2)_{D}\ltimes\mathbb{R}^{3}$
generated by (\ref{LsD})-(\ref{QsD}), the bulk magnon transforms in some tensor product representation
$\boxslash_{(a',b',c',d')}\otimes
{\boxslash}_{(\tilde{a},\tilde{b},\tilde{c},\tilde{d})}\,,$
where $(a',b',c',d')$ and $(\tilde a, \tilde b, \tilde c, \tilde d)$ are representation labels that we must determine. For the left factor we have simply $(a',b',c',d')=(a,b,c,d)$. For the right factor, the choices of defect orientation and gamma matrices made in \cite{Correa1} lead to the relations $\bar{\dot{3}}=\dot{4}$ and $\bar{\dot{4}}=\dot{3}$. Because of this and the $\kappa$ factors appearing in (\ref{QsD}), one must change basis in order for the action of the generators (\ref{LsD})-(\ref{QsD}) to be the canonical one: 
\begin{equation}
(\tilde{\phi}^{\check{1}},\tilde{\phi}^{\check{2}}|\tilde{\psi}^{\check{3}},\tilde{\psi}^{\check{4}}):=
(\tilde{\phi}^{\dot{1}},\tilde{\phi}^{\dot{2}}|\kappa\tilde{\psi}^{\dot{4}},\kappa\tilde{\psi}^{\dot{3}})\,. \label{newbasis}
\end{equation}
One then finds that $(\tilde{a},\tilde{b},\tilde{c},\tilde{d})=(a,-\kappa^{2}b,-\kappa^{-2}c,d)$.
Therefore, with respect to the boundary symmetry algebra $\mathfrak{psu}(2|2)_{D}\ltimes\mathbb{R}^{3}$
generated by (\ref{LsD})-(\ref{QsD}), the bulk magnon transforms in the representation
\begin{equation}
{\boxslash}_{(a,b,c,d)}\otimes
{\boxslash}_{(a,-\kappa^{2}b,-\kappa^{-2}c,d)}.\label{tensorb}
\end{equation}

Representations of $\mathfrak{psu}(2|2)_{D}\ltimes\mathbb{R}^{3}$ can also be labelled by the values of the central charges \cite{Beis2006}: for the fundamental representation we have $\boxslash_{(a,b,c,d)} \cong \langle 0,0;H,C,C^\dagger \rangle$ where $H=ad+bc$, $C=ab$, $C^\dagger=cd$.  In these terms, the bulk magnon lives in the representation
\begin{equation}
\left\langle 0,0;H,C,C^{\dagger}\right\rangle \otimes\left\langle 0,0;H,-\kappa^{2}C,-\kappa^{-2}C^{\dagger}\right\rangle
\cong\left\{ 0,0,2H,(1-\kappa^{2})C,(1-\kappa^{-2})C^{\dagger}\right\}.\label{ccsb}
\end{equation}
We now consider separately the horizontal and vertical vacua.

\begin{figure}
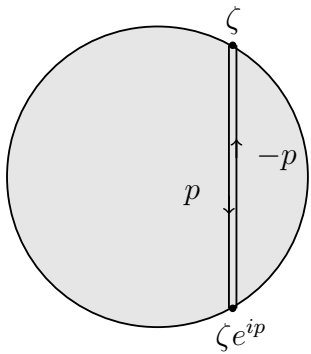
 \begin{center}\btp
    \filldraw[fill=gray!20!white] (8,8) circle (8);
    \filldraw (12,1) circle (1.5mm);
    \filldraw (12,15) circle (1.5mm);
    \draw[-] (11.8,1) -- (11.8,15) node [above] {$\;\zeta$};
    \draw[-] (12.2,15) -- (12.2,1) node [below] {$\;\zeta e^{ip}$};
    \draw[->] (11.8,7) node [left] {$p\;\;$} -- (11.8,6);
    \draw[->] (12.2,9) node [right] {$\;-p$} -- (12.2,10);
\etp
\end{center}
\hspace{1in}\parbox{4in}{\caption{Bulk magnon diagonal representation
looks like two consecutive magnons with momentum $p$ and phase $\zeta$ and
momentum $-p$ and phase $\zeta e^{ip}$.} \label{doublemag}}
\end{figure}

\paragraph{Horizontal vacuum.} This case corresponds to $\kappa=-1$
in (\ref{QsD}).
With respect to $\mathfrak{psu}(2|2)_{D}\ltimes\mathbb{R}^{3}$, the bulk magnon transforms in the tensor representation
\begin{equation}
\boxslash_{(a,b,c,d)}\otimes \boxslash_{(a,-b,-c,d)} = \mathcal{V}\left(p,\zeta\right) \otimes \mathcal{V}\left(-p,\zeta e^{ip}\right)\label{(p,-p) rep}\,.
\end{equation}
One way of interpreting this is as two consecutive magnons with momenta $p$ and $-p$, as depicted in figure \ref{doublemag}.
The central
charges $\check{\mathbb{C}}$ and $\mathbb{C}^{\dagger}$ vanish and
\begin{equation}
\left\langle 0,0;H,C,C^{\dagger}\right\rangle \otimes\left\langle 0,0;H,-C,-C^{\dagger}\right\rangle =\left\{ 0,0,2H,0,0\right\} .
\end{equation}

\paragraph{Vertical vacuum.} This case corresponds to $\kappa=-i$.
The bulk magnon transforms in the tensor representation
\begin{equation}
\boxslash_{(a,b,c,d)}\otimes\boxslash_{(a,b,c,d)}= \mathcal V(p,\zeta) \otimes \mathcal V(p,\zeta)\,,\label{vvrep}
\end{equation}
under $\mathfrak{psu}(2|2)_{D}\ltimes\mathbb{R}^{3}$. The values of the central charges $\check{\mathbb C}$, $\check{\mathbb C}^{\dagger}$ and $\check{\mathbb H}$ are
\begin{equation}
\check{H}=2H,\quad\check{C}=2C,\quad\check{C}^{\dagger}=2C,
\end{equation}
These values satisfy the multiplet splitting condition
$\check{H}^{2}-\check{C}\check{C}^{\dagger}=1$, according to which
\begin{equation}
\left\{ 0,0;2H,2C,2C^{\dagger}\right\} =\left\langle 1,0;2H,2C,2C^{\dagger}\right\rangle \oplus\left\langle 0,1;2H,2C,2C^{\dagger}\right\rangle = \twobh \oplus \twobv.
 \end{equation}
Thus, bulk magnons transform in the direct sum of symmetric and antisymmetric short representations of $\mathfrak{psu}(2|2)_{D}\ltimes\mathbb{R}^{3}$. In the conventions of \cite{Arutyunov1,Arutyunov2} these representations are equivalent to two-magnon ($M=2$) bound-state and mirror bound-state representations with the labels having doubled
coupling constant, i.e. $g$ replaced by $2g$. This doubling of the coupling constant dependence of the labels of the diagonal representations explains why the all-loop dispersion relation is still given by $H(p) = \sqrt{1+8g^2\sin^2(\tfrac{p}{2})}$ in spite of having $M=2$
reps\footnote{For the $M=2$ reps we have $\check H =2H = \sqrt{2^2+8(2g)^2\sin^2(\tfrac{p}{2})}$.}. We shall observe a similar duplication for the boundary representation labels in this case.

\subsection{Reflection matrix: horizontal case}\label{sshorz}
We consider first the reflection of a bulk magnon from the boundary in the horizontal case. For definiteness, we consider a right boundary. In the horizontal case the boundary is a singlet. Then reflection from a right boundary sends $p\mapsto -p$ and $\zeta\mapsto\zeta$ \cite{HMopen}.
Thus the reflection matrix is a map
\begin{equation}
\mathcal{K}^{h}:\quad\mathcal{V}\left(p,\zeta\right)\otimes\mathcal{V}(-p,\zeta e^{ip})\otimes 1 \rightarrow\mathcal{V}(-p,\zeta)\otimes\mathcal{V}(p,\zeta e^{-ip})\otimes 1\,.
\end{equation}
As noted above, the tensor representation (\ref{(p,-p) rep}) corresponds to two consecutive magnons with momenta $p$ and $-p$. Therefore the $K$-matrix $\mathcal{K}^h(p,-p)$ intertwines the same representations as the bulk $S$-matrix $\mathcal{S}(p,-p)$, and the two must be equal up to a phase (since this intertwiner is fixed by symmetry, up to a phase).
Details of $\mathcal{S}$ and $\mathcal{K}^h$ are given in appendices A and B.

In order to check that the boundary is integrable one has to consider the
boundary Yang-Baxter equation (bYBE), which computes the difference between
the two possible ways of factorizing the scattering of two incoming
magnons off a boundary.

It is convenient to do all the calculations in terms of representations of the preserved boundary symmetry algebra $\mathfrak{psu}(2|2)_{D}\ltimes\mathbb{R}^{3}$ generated by (\ref{LsD})-(\ref{QsD}).
The bYBE represents two incoming bulk magnons reflecting from the boundary:
\begin{align}
 & \mbox{bYBE}:\mathcal{V}_{L}(p_{1},\zeta)\otimes\mathcal{V}_{R}(-p_{1},\zeta e^{ip_1})\otimes
    \mathcal{V}_{L}(p_{2},\zeta e^{ip_{1}})\otimes\mathcal{V}_{R}(-p_{2},\zeta e^{i(p_1+p_2)})\rightarrow \nonumber \\
 & \qquad\qquad\mathcal{V}_{L}(-p_{1},\zeta)\otimes\mathcal{V}_{R}(p_{1},\zeta e^{-ip_1})\otimes
    \mathcal{V}_{L}(-p_{2},\zeta e^{-ip_{1}})\otimes\mathcal{V}_{R}(p_{2},\zeta e^{-i(p_1+p_2)}).\label{bYBE_h}
\end{align}
Here $\mathcal{V}_{L}$ ($\mathcal{V}_{R}$) are representations of the boundary algebra originating as left (respectively, right) factors of bulk magnons. We must not lose track of this information, because it affects how the representations scatter, as follows.

For the bulk scattering, left (right) states scatter with left (respectively, right) states only. When scattering two left representations we use the standard $S$-matrix, but when scattering two right representations we must allow for the change of basis, (\ref{newbasis}), which produces additional signs in the $\zeta$-dependent components:
\begin{align}
 & \langle \psi^{\check{3}}\psi^{\check{4}} | \,\mathcal{S}\, | \phi^{\check{1}}\phi^{\check{2}} \rangle
  =
 - \langle \psi^{\dot{3}}\psi^{\dot{4}} | \,\mathcal{S}\, | \phi^{\dot{1}}\phi^{\dot{2}} \rangle
 = - a_7
  \nonumber \\
 & \langle \phi^{\check{1}}\phi^{\check{2}} | \,\mathcal{S}\, | \psi^{\check{3}}\psi^{\check{4}} \rangle
  =
 - \langle \phi^{\dot{1}}\phi^{\dot{2}} | \,\mathcal{S}\, | \psi^{\dot{3}}\psi^{\dot{4}} \rangle
 = - a_8.
  \label{signchange}
\end{align}
Given that $a_7$ and $a_8$ depend linearly on the phase, this sign change is just $\zeta \mapsto -\zeta$.
Next, to exchange a left state with a right state in the tensor product one must use a graded permutation, which also produces certain minus signs.\footnote{This graded permutation was overlooked in the calculations of \cite{Correa1} thus obscuring the integrability of the $D$5-brane boundary conditions.}

\begin{figure}[ht]
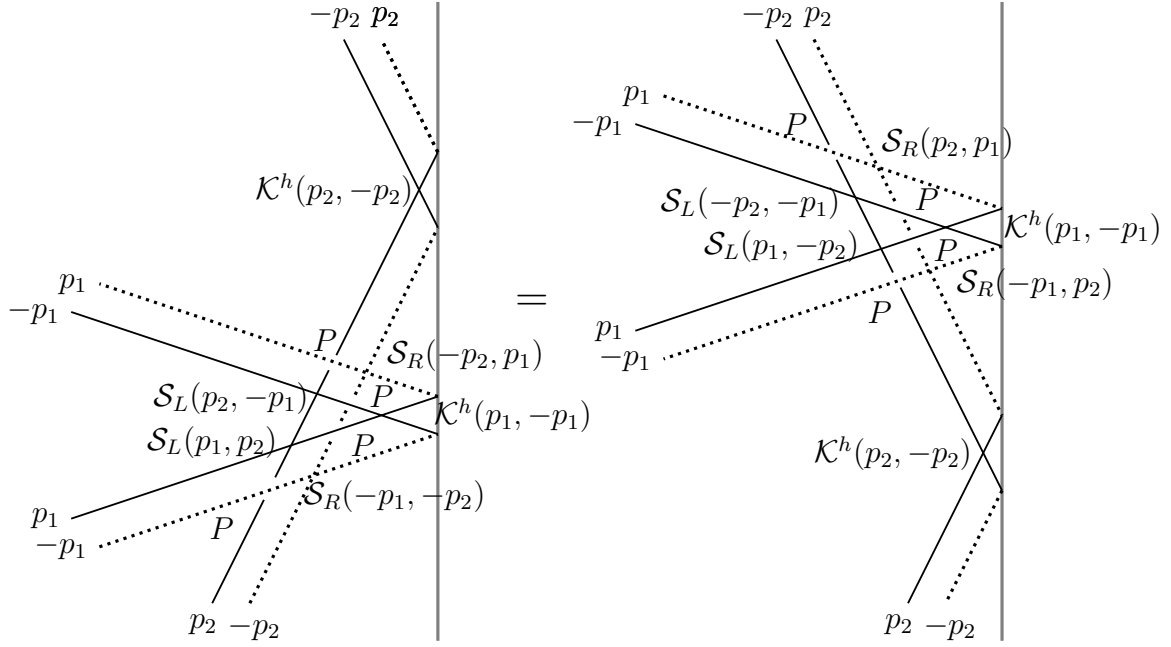
 \begin{center} \btp
    \draw[gray,very thick,-] (20,-2) -- (20,32);
    \draw[-] (8,0) node [below] {$p_2\;\;$} -- (10.75,5.5); 
    \draw[-]  (11.35,6.7) -- (14.2,12.4);
    \draw[-]  (14.6,13.2) -- (19,22) -- (15,30) node [above] {$-p_2\;\;$};
    \draw[-,very thick,dotted] (10,0) node [below] {$\;-p_2$} -- (14.4,8.8);
    \draw[-,very thick,dotted] (14.85,9.7) -- (15.2,10.4);
    \draw[-,very thick,dotted] (15.6,11.1) -- (20,20);
    \draw[-,very thick,dotted]  (20,24) -- (17,30) node [above] {$\;p_2$};
    \draw[-](20,20) -- (19,22) -- (20,24);
    \draw[-,very thick,dotted] (20,24) -- (17,30) node [above] {$\;p_2$};
    \draw[-] (0.5,4.5) node [left] {$p_1$} -- (17,10) -- (0.5,15.5) node [left] {$-p_1$};
    \draw[-,very thick,dotted] (2,3) node [left] {$-p_1$} -- (20,9);
    \draw[-] (20,9) -- (17,10) -- (20,11);
    \draw[-,very thick,dotted] (20,11) -- (2,17) node [left] {$p_1$};
    \draw (17.7,5.7) node {$\mathcal{S}_R(-p_1,-p_2)$} (16.7,13.2) node [right] {$\mathcal{S}_R(-p_2,p_1)$};
    \draw (8,8.6) node {$\mathcal{S}_L(p_1,p_2)$} (9,10.9) node {$\mathcal{S}_L(p_2,-p_1)$};
    \draw (24,10) node {$\mathcal{K}^h(p_1,-p_1)$} (14.5,22) node {$\mathcal{K}^h(p_2,-p_2)$} (14,14) node {$P$} (8.5,4) node {$P$} (16,8.4) node {$P$} (17,11) node {$P$};

    \draw[gray,very thick,-] (50,-2) -- (50,32);
    \draw[-] (38,30) node [above] {$-p_2\;\;$} -- (40.8,24.4);
    \draw[-] (41.2,23.6) -- (44.2,17.6);
    \draw[-] (44.6,16.8) -- (49,8) -- (45,0) node [below] {$p_2\;\;$};
    \draw[-,very thick,dotted] (40,30) node [above] {$\;p_2$} -- (44.4,21.2);
    \draw[-,very thick,dotted] (44.8,20.4)  -- (45.2,19.6);
    \draw[-,very thick,dotted] (45.55,18.9)  -- (50,10);
    \draw[-] (50,10) -- (49,8) -- (50,6);
    \draw[-,very thick,dotted] (50,6) -- (47,0) node [below] {$\;-p_2$};
    \draw[-] (30.5,25.5) node [left] {$-p_1$} -- (47,20) -- (30.5,14.5) node [left] {$p_1$};
    \draw[-,very thick,dotted] (32,27) node [left] {$p_1$} -- (50,21);
    \draw[-] (50,21) -- (47,20) -- (50,19);
    \draw[-,very thick,dotted] (50,19) -- (32,13) node [left] {$-p_1$};
    \draw (49,8) node [left] {$\mathcal{K}^h(p_2,-p_2)$} (49.5,20) node [right] {$\mathcal{K}^h(p_1,-p_1)$} (43.5,15.5) node {$P$} (47,18.8) node {$P$} (39,25.5) node {$P$} (46,21.5) node {$P$};
    \draw (47,17) node [right] {$\mathcal{S}_R(-p_1,p_2)$} (43,24.5) node [right] {$\mathcal{S}_R(p_2,p_1)$};
    \draw (43,19) node [left] {$\mathcal{S}_L(p_1,-p_2)$} (42,21.3) node [left] {$\mathcal{S}_L(-p_2,-p_1)$};

    \draw[black,thick] (25,16) node {\Large{=}};
\etp \end{center}
\hspace{1in}\parbox{4in}{\caption{bYBE for the reflection in the horizontal case.
Solid lines correspond to the left reps while the dotted lines correspond to right reps.
} \label{fig_bYBE}}
\end{figure}

The pictorial version of the bYBE is presented in figure \ref{fig_bYBE} and the equation
itself is
\begin{align}
 & \mathcal{K}_{34}(p_{2},\zeta e^{-ip_{1}};-p_{2},\zeta e^{-i(p_{1}-p_{2})})\,P_{23}\, \mathcal{S}_{34}(-p_{2},-\zeta e^{ip_{2}};p_{1},-\zeta e^{i(p_{2}-p_{1})}) \nonumber \\
 & \qquad\times \mathcal{S}_{12}(p_{2},\zeta;-p_{1},\zeta e^{ip_{2}})\, P_{23}\, \mathcal{K}_{34}(p_{1},\zeta e^{ip_{2}};-p_{1},\zeta e^{i(p_{1}+p_{2})}) \nonumber \\
 & \qquad\qquad\times P_{23}\, \mathcal{S}_{12}(p_{1},\zeta;p_{2},\zeta e^{ip_{1}})\, \mathcal{S}_{34}(-p_{1},-\zeta e^{ip_{1}};-p_{2},-\zeta e^{i(p_{1}+p_{2})})\, P_{23} \nonumber \\
 & -P_{23}\, \mathcal{S}_{34}(p_{2},-\zeta e^{-ip_{2}};p_{1},-\zeta e^{-i(p_{2}+p_{1})})\, \mathcal{S}_{12}(-p_{2},\zeta;-p_{1},\zeta e^{-ip_{2}})\, P_{23} \nonumber \\
 & \qquad\times \mathcal{K}_{34}(p_{1},\zeta e^{-p_{2}};-p_{1},\zeta e^{-i(p_{2}-p_{1})}) \, P_{23}\, \mathcal{S}_{12}(p_{1},\zeta;-p_{2},\zeta e^{ip_{1}}) \nonumber \\
 & \qquad\qquad\times  \mathcal{S}_{34}(-p_{1},-\zeta e^{ip_{1}};p_{2},-\zeta e^{i(p_{1}-p_{2})})\, P_{23} \, \mathcal{K}_{34}(p_{2},\zeta e^{ip_{1}};-p_{2},\zeta e^{i(p_{1}+p_{2})}) =0,\label{bYBE_D_alg}
\end{align}
where the subscripts $12$, $23$, $34$ indicate the tensor factors on which the operators act, $P_{ij}$ is the graded permutation operator permuting left-right states, $\mathcal{S}_{ij}^{L}$ and $\mathcal{S}_{ij}^{R}$ are the left and right bulk $S$-matrices and $\mathcal{K}_{34}$ is the reflection matrix. We have checked directly that this boundary YBE is satisfied.

\begin{figure}[ht]
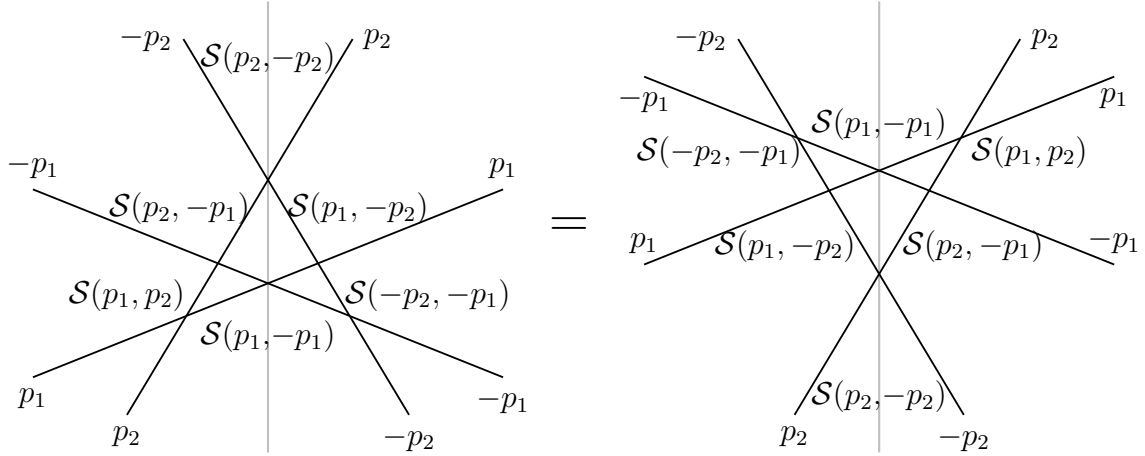
 \begin{center} \btp
    \draw[lightgray,thick,-] (12.5,-2) -- (12.5,22);
    \draw[black] (5,0) node [below] {$p_2$} -- (17,20) node [right] {$p_2$};
    \draw[black] (8,20) node [left] {$-p_2$} -- (20,0) node [below] {$-p_2$};
    \draw[black] (0,2) node [below] {$p_1$} -- (25,12) node [above] {$p_1$};
    \draw[black] (0,12) node [above] {$-p_1$} -- (25,2) node [below] {$-p_1$};
    \draw (5,6.4) node {$\mathcal{S}(p_1,p_2)$} (21,6.4) node {$\mathcal{S}(-p_2,-p_1)$} (12.5,4.2) node {$\mathcal{S}(p_1,\!-p_1)$}
	  (12.5,19) node {$\mathcal{S}(p_2,\!-p_2)$} (7.8,11) node {$\mathcal{S}(p_2,-p_1)$} (17.3,11) node {$\mathcal{S}(p_1,-p_2)$};

    \draw[lightgray,thick,-] (45,-2) -- (45,22);
    \draw[black] (37.5,20) node [left] {$-p_2$} -- (49.5,0) node [below] {$-p_2$};
    \draw[black] (40.5,0) node [below] {$p_2$} -- (52.5,20) node [right] {$p_2$};
    \draw[black] (32.5,18) node [below] {$-p_1$} -- (57.5,8) node [above] {$-p_1$};
    \draw[black] (32.5,8) node [above] {$p_1$} -- (57.5,18) node [below] {$p_1$};
    \draw[black] (40,9) node {$\mathcal{S}(p_1,-p_2)$} (50,9) node {$\mathcal{S}(p_2,-p_1)$} (45,15.5) node {$\mathcal{S}(p_1,\!-p_1)$}
	  (45,1) node {$\mathcal{S}(p_2,\!-p_2)$} (36.5,14) node {$\mathcal{S}(-p_2,-p_1)$} (53,14) node {$\mathcal{S}(p_1,p_2)$}
	  (28.5,10) node {\LARGE{=}};
\etp \end{center}
\caption{The unfolded bYBE as a 4-particle scattering in the bulk. The vertical line plays no role in the unfolded picture, but is drawn as a reminder that  $S(p,-p)$  represents boundary reflections.}\label{bYBE_unfolded}
\end{figure}

Another way to verify that the boundary YBE is satisfied is to note that it may be mapped to a standard \emph{bulk} YBE, as follows.  One can verify that whenever a phase-dependent component appears, the extra sign in the right $S$-matrix is canceled with a minus sign from a graded permutation. Then, using also the relation between $\mathcal{K}$ and the bulk S-matrix, the above equation is equivalent to
\begin{align}
& \mathcal{S}_{23}(p_{2},\zeta e^{-ip_{1}};-p_{2},\zeta e^{ip_{2}-ip_{1}})\,\mathcal{S}_{34}(p_{1},\zeta e^{ip_{2}-ip_{1}};-p_{2},\zeta e^{ip_{2}})\,\mathcal{S}_{12}(p_{2},\zeta;-p_{1},\zeta e^{ip_{2}})\nonumber\\
& \qquad\times\mathcal{S}_{23}(p_{1},\zeta e^{ip_{2}};-p_{1},\zeta e^{ip_{1}+ip_{2}})\,\mathcal{S}_{34}(-p_{2},\zeta e^{ip_{1}+ip_{2}};-p_{1},\zeta e^{ip_{1}})\,\mathcal{S}_{12}(p_{1},\zeta;p_{2},\zeta e^{ip_{1}})\nonumber\\
& -\mathcal{S}_{34}(p_{1},\zeta^{-ip_{1}-ip_{2}};p_{2},\zeta e^{-ip_{2}})\,\mathcal{S}_{12}(-p_{2},\zeta;-p_{1},\zeta e^{-ip_{2}})\,\mathcal{S}_{23}(p_{1},\zeta e^{-ip_{2}};-p_{1},\zeta e^{ip_{1}-ip_{2}})\nonumber\\
& \qquad\times\mathcal{S}_{34}(p_{2},\zeta e^{ip_{1}-ip_{2}};-p_{1},\zeta e^{ip_{1}})\,\mathcal{S}_{12}(p_{1},\zeta;-p_{2},\zeta e^{ip_{1}})\,\mathcal{S}_{23}(p_{2},\zeta e^{ip_{1}};-p_{2},\zeta e^{ip_{1}+ip_{2}}) =0.
\label{bYB Eq_unfolded}
\end{align}
In this way we have ``unfolded'' the bYBE into a succession of bulk scattering
processes.  Consequently, the boundary YBE follows from a particular case of the bulk YBE. The meaning of (\ref{bYB Eq_unfolded}) is represented in figure \ref{bYBE_unfolded}.

\noindent
From this second, ``unfolded'', point of view, the boundary is seen to be ``achiral'', meaning that an incoming left state becomes a right one after the reflection and a right one becomes a left.

\subsection{Reflection matrix: vertical case}
In the vertical case the boundary carries a degree of freedom
transforming in a fundamental representation $\boxslash_{(a_B,b_B,c_B,d_B)}$
of $\mathfrak{psu}(2|2)_{D}\ltimes\mathbb{R}^{3}$. The representation labels specifying this representation
are\footnote{Note that here we use a different parametrization from the one used in \cite{Correa1}.}
\begin{equation}
a_{B}=\sqrt{g}\eta_{B},\quad b_{B}=-\sqrt{g}\frac{i\zeta}{\eta_{B}},\quad c_{B}=-\sqrt{g}\frac{\eta_{B}}{\zeta x_{B}},\quad d_{B}=\sqrt{g}\frac{x_{B}}{i\eta_{B}}.\label{abcd_B}
\end{equation}
This representation is related to a radial line segment in the LLM disc
picture \cite{HMopen,HM,LLM}.
The unitarity and mass-shell conditions give
\begin{equation}
|\eta_{B}|^{2}=-ix_{B}\,,\qquad x_{B}\equiv\frac{i(1+\sqrt{1+4g^{2}})}{2g}\,.
\label{mass-shell_B}
\end{equation}
Thus, the exact energy of the boundary excitation is
\begin{equation}
\check{H}=\mathcal{D}-J_{56}=\frac{1}{2}\sqrt{1+4g^{2}}\,.\label{exacte}
\end{equation}
For the boundary degree of freedom, the representation labels (\ref{abcd_B})
and the mass-shell condition (\ref{mass-shell_B}) are those of the boundary fundamental degree of freedom in the D3-brane case \cite{HMopen}, but with a coupling constant $g$  twice bigger. This doubling of the coupling constant is crucial for integrability to hold and for the exact boundary energy (\ref{exacte}) to consistently reproduce 1-loop anomalous dimensions.

As we saw, the elementary bulk magnons transform, under the boundary symmetry algebra, in direct sum of two $M=2$ bound state representations (symmetric and antisymmetric).
Therefore we have the following two scattering processes:
\begin{align}
 & \mathcal{K}:\twobh\otimes\,\boxslash\rightarrow\twobh\otimes\,\boxslash,\label{SymRef}\\
 & \overline{\mathcal{K}}:\twobv \otimes \boxslash \rightarrow \twobv \otimes \boxslash.\label{AsymRef}
\end{align}
As in \cite{Correa1,MR1}, following \cite{Arutyunov1}, the reflection matrices in the symmetric and antisymmetric channels (\ref{SymRef}) are
\begin{equation}
\mathcal{K}^{Ba}=\sum_{i=1}^{19}k^{\text{\tiny (S)}}_{i}\,\Lambda_{i},\qquad \overline{\mathcal{K}}^{Ba}=\sum_{i=1}^{19} k^{\text{\tiny (A)}}_{i}\bar{\Lambda}_{i},
\end{equation}
where $\Lambda_{i}$ are certain differential operators (see appendix C for details), $\bar{\Lambda}_{i}$ are obtained from $\Lambda_{i}$ by exchanging indices $\check{1}\leftrightarrow\check{3}$ and $\check{2}\leftrightarrow\check{4}$, and where $k^{\text{\tiny (S,A)}}_{i}$ are the reflection coefficients.
In both cases, the symmetry algebra alone fixes all reflection coefficients up to an overall phase. Interestingly, the two channels are related by
$k^{\text{\tiny (A)}}_{i}(p,x_B)=k^{\text{\tiny (S)}}_{i}(-p,x_B)$ \cite{Correa1}. Note that the reflection coefficients do not explicitly depend on $g$, thus they coincide with the ones found in \cite{MR1}.

It is easy to check that symmetric and antisymmetric reflection matrices
$\mathcal{K}^{Ba}$ and $\overline{\mathcal{K}}^{Ba}$ do satisfy bYBE on their own. The
bYBE invariance of $\mathcal{K}^{Ba}$ was checked in \cite{MR1}, while
for checking the bYBE invariance of $\overline{\mathcal{K}}^{Ba}$  we had to
construct an antisymmetric bound state $S$-matrix $\overline{\mathcal{S}}^{BB}$
which is the mirror-model partner of the ordinary bound state $S$-matrix
$\mathcal{S}^{BB}$.

For the vertical vacuum case the complete  reflection matrix must be some linear combination:
\begin{equation}
\mathcal{K}^{v}=k_{0}\,\mathcal{K}^{Ba}+\,\overline{\mathcal{K}}^{Ba}, \label{KV}
\end{equation}
with $k_{0}$ being a function of bulk and boundary representation parameters.
The important question is whether there exists any choice of this
function, such that the system is integrable, i.e. such that the
complete reflection matrix obeys the boundary Yang-Baxter equation.
For this purpose one needs to consider the complete bulk $16\times16$-dim.
$S$-matrix $\mathcal{S}^{A\dot{A}A\dot{A}}$ which may be constructed as a
tensor product of two fundamental $S$-matrices $\mathcal{S}^{AA}$ and $\mathcal{S}^{\dot{A}\dot{A}}$.
It is convenient to compute $\mathcal{S}^{A\dot{A}A\dot{A}}$ in the basis of
(graded) symmetric and antisymmetric states i.e. on the superspace and the
mirror-superspace. The complete $S$-matrix is not block-diagonal in this basis; rather it mixes symmetric and antisymmetric states during the scattering. 
But it is important to note that it is invariant
under the symmetries preserved by the boundary (this is natural as
the boundary algebra is a subalgebra of the bulk algebra).

The bYBE for the reflection in this vertical case reads as
\begin{align}
 & {\rm bYBE}:\mathcal{V}_{L}(p_{1},\zeta) \otimes \mathcal{V}_{R}(p_{1},\zeta) \otimes
    \mathcal{V}_{L}(p_{2},\zeta e^{ip_{1}}) \otimes \mathcal{V}_{R}(p_{2},\zeta e^{ip_1}) \otimes \mathcal{V}_{B}(x_B,\zeta e^{i(p_1+p_2)}) \rightarrow \nonumber \\
 & \qquad \mathcal{V}_{L}(-p_{1},\zeta) \otimes \mathcal{V}_{R}(-p_{1},\zeta) \otimes
    \mathcal{V}_{L}(-p_{2},\zeta e^{-ip_{1}}) \otimes \mathcal{V}_{R}(-p_{2},\zeta e^{-ip_1}) \otimes \mathcal{V}_{B}(x_B,\zeta e^{-i(p_1+p_2)}),\label{bYBE_v}
\end{align}
where once again the scattering in the bulk is between left-left and right-right
states only, while the permutation of left-right and right-left states produces
a graded minus sign. In the contrast to the horizontal case, the right $S$-matrix is equivalent to the left $S$-matrix, i.e. it does not acquire an extra minus sign in the  $\zeta$-dependent components, since now $-\kappa^{2}=+1$. Also, all phases in (\ref{bYBE_v}) are increasing from left to right. The graphical interpretation of
bYBE is almost the same as for the horizontal case. The difference is
that the boundary in this case does not act diagonally but mixes bulk and boundary flavours.

A general matrix element of the bYBE (\ref{bYBE_v}) has a complicated structure.
We found the particular matrix element
\begin{equation}
\left\langle \phi_{1}^{\{\check{3}\check{4}\}}\otimes\phi_{2}^{\{\check{1}\check{1}\}} \otimes
  \phi_{B}^{\check{1}}\right| {\rm bYBE} \left|\psi_{1}^{\{\check{1}\check{3}\}} \otimes
  \phi_{2}^{\{\check{1}\check{1}\}}\otimes\psi_{B}^{\check{4}}\right\rangle
  \label{bybek0}
\end{equation}
to be quite tractable and by treating minus signs coming from permuting
left and right reps carefully (i.e. {\small $\mathcal{S}^{A_{1}\dot{A}_{2}A_{3}\dot{A}_{4}} =
(-1)^{[\dot{A}_{2}][A_{3}]}\mathcal{S}^{A_{1}A_{3}} \otimes \mathcal{S}^{\dot{A}_{2}\dot{A}_{4}}$})
we find the required ratio has to be
\begin{align}
k_{0}=-\frac{x^{-}(x_{B}-x^{-})^{2}}{x^{+}(x_{B}+x^{+})^{2}}
\frac{\eta^{2}\eta_{B}}{\tilde{\eta}^{2}\tilde{\eta}_{B}}.
\label{k0}
\end{align}
for (\ref{bybek0}) to vanish. We have then checked that, using this ratio, the reflection matrix $K^v$ (\ref{KV}) satisfies all matrix elements of bYBE (\ref{bYBE_v}). Thus we conclude that the reflection in the vertical case is indeed integrable. We also claim that it is an achiral boundary in the same sense as in the horizontal case: at this stage the ``unfolded'' picture of the reflection is not obvious, but it will become clear when we consider the nested Bethe ansatz.

To end this section we would like to produce  a weak coupling consistency check for the ratio $k_0$ between the reflection of symmetric and antisymmetric components. We shall focus on the right reflection of components $\phi^{\{11\}}$ of $\boxslash\hspace{-0.9mm}\boxslash$ and $\phi^{[12]}$ of $\twobv$, both of which are reflected diagonally when the right boundary has the defect field $\phi^1$. Using the exact expressions for (\ref{k0}) and (\ref{ks}), we expand in powers of $g^2$ and obtain\footnote{We consider $\eta$'s in the spin chain basis, in which $\frac{\eta^{2}\eta_{B}}{\tilde{\eta}^{2}\tilde{\eta}_{B}}=1$.}
\begin{equation}
\frac{{K }_{R}^{\phi^{\{11\}}}}{{ K}_{R}^{\phi^{[12]}}} =
\frac{k_{0} k_{1}^{\text{\tiny (S)}}}{k_{9}^{\text{\tiny (A)}}}
=
\frac{3-e^{ip}}{1-3 e^{ip}}+ {\cal O}(g^2)\, .
\end{equation}
The leading order is in exact agreement with the ratio
\begin{equation}
\frac{{K}^{Y}_L(-p)}{{K}^{X_V^1}_L(-p)} =
\frac{3-e^{ip}}{1-3 e^{ip}}\,,
\end{equation} of the reflection factors obtained in the appendix \ref{1lopp} from the 1-loop mixing matrix of anomalous dimensions. Therefore, the unique choice of $k_0$ consistent with the bYBE is also consistent with the available weak coupling results.


\section{Coordinate Bethe Ansatz}
\label{CBA}

We now proceed to derive the asymptotic Bethe equations for the vertical case, by means of the nested coordinate Bethe ansatz \cite{yang}. Our treatment is similar to that in  \cite{Correa2}.

\subsection{Bethe ansatz}

Let us start by considering the scattering problem on the half-line
with a right boundary. As in the previous section, we work in terms of representations of the symmetry algebra $\mathfrak{psu}(2|2)_{D}\ltimes\mathbb{R}^{3}$ preserved by the boundary.
An asymptotic state with $N^{\rm I}$ elementary bulk magnons of momenta $(p_1,\dots,p_{N^{\rm I}})$ and phases $(\zeta_1,\dots,\zeta_{N^{\rm I}})$ then transforms in the representation (c.f. \ref{vvrep})
\begin{equation}
\Big(\mathcal V(p_1,\zeta_1)\otimes\mathcal V(p_1,\zeta_1)\Big)\otimes\dots\otimes
\Big(\mathcal V(p_{N^{\rm I}},\zeta_{N^{\rm I}})\otimes\mathcal V(p_{N^{\rm I}},\zeta_{N^{\rm I}})\Big)\otimes
\mathcal V_B(\zeta_B)
\label{asym_state}
\end{equation}
with $2N^{\rm I}+1$ tensor factors.
We would like to go to an ``unfolded'' picture of the boundary, as we did for the horizontal case in \S\ref{sshorz}, so let us choose to write these tensor factors in a different order, namely
\begin{align}
\mathcal V(p_1,\zeta_1) \otimes \dots \mathcal V(p_{N^{\rm I}},\zeta_{N^{\rm I}}) \otimes \mathcal V_B(\zeta_B) \otimes  \mathcal V(p_{N^{\rm I}},\zeta_{N^{\rm I}}) \otimes \dots \otimes \mathcal V(p_1,\zeta_1).\label{asym_state_unfold}
\end{align}
Doing so introduces minus signs when permuting fermions, which we need to keep careful account of below.
We shall write basis vectors as
\begin{equation}
\left|\chi_{1}^{a}\ldots\chi_{N^{\rm I}}^{y}\chi_{B}^{z}\chi_{N^{\rm I}}^{y}\ldots\chi_{1}^{a}\right\rangle,\label{basis_state}
\end{equation}
where all indices run over $1\ldots4$.  We no longer decorate indices that originated as right multiplets with dots, nor the boundary ones with checks. After all, these indices all transform canonically under the preserved symmetry algebra. (So, $a$ should strictly be $\bar{a}$ in the notation of section 2.)

We write $\mathcal{S}_{i,i+1}^{\rm I}$ for the fundamental left or right
$S$-matrix. As noted above, these are identical since $-\kappa^{2}=+1$. We write $\mathcal{K}^{\rm I}$ for the reflection matrix found in (\ref{KV}).

\subsubsection*{Level II}

We start from defining the \textit{level II} vacuum to be the state
consisting only of $\psi^{3}$
\begin{equation}
\left|0\right\rangle ^{\! \rm II}=\left|\psi_{1}^{3}\ldots\psi_{N^{\rm I}}^{3}\psi_{B}^{3}\psi_{N^{\rm I}}^{3}\ldots\psi_{1}^{3}\right\rangle ,
\end{equation}
On this state $\mathcal{S}_{i,i+1}^{\rm I}$ and $\mathcal{K}^{\rm I}$ act diagonally: $\mathcal{S}_{i,i+1}^{\rm I}\left| 0 \right>^{\! \rm II} = \left| 0 \right>^{\! \rm II} S_{i,i+1}^{\mathrm I}$, $\mathcal{K}^{\rm I}\left| 0 \right>^{\! \rm II} = \left| 0 \right>^{\! \rm II} K^{\mathrm I}$.   We normalize the scattering and reflection matrices in such way that
\begin{equation}
S_{i,i+1}^{\rm I}=-1,\qquad\mbox{and}\qquad K^{\rm I}=+1.
\label{levelI}
\end{equation}

\paragraph{Reflection of level II excitations.} Next we define \textit{level II} excitations, defined to transform under $\mathcal{S}_{i,i+1}^{\rm I}$ and $\mathcal{K}^{\rm I}$ in exactly the same fashion as $\left|0\right\rangle ^{\! \rm II}$ (the \emph{compatibility condition}).

We consider first a single excitation. As usual we make a spin-wave ansatz in which the particle has a ``tail'' running away behind it. This ansatz is the sum of an ``ingoing'' and an ``outgoing'' piece, plus a term in which the excitation has just reached the boundary.  So we have, using a pictorial notation,
\begin{align}
\left|\Psi^{a}(y)\right\rangle^{\! \rm II} & =\ldots + \btp
	\draw[dotted] (5,-2) -- (5,2);
	\draw[dotted] (4,-2) -- (4,2);
        \draw[black,line width=1.5pt] (3,-2) -- (3,2);
        \draw[dotted] (2,-2) -- (2,2);
        \draw[dotted] (1,-2) -- (1,2);
        \draw[->] (0,-2)   -- (2,1);\etp \,\,\,
+ \btp
	\draw[dotted] (5,-2) -- (5,2);
	\draw[dotted] (4,-2) -- (4,2);
        \draw[black,line width=1.5pt] (3,-2) -- (3,2);
        \draw[dotted] (2,-2) -- (2,2);
        \draw[dotted] (1,-2) -- (1,2);
        \draw[->] (0,-2)   -- (3,1);\etp \,\,\,
+ \btp
	\draw[dotted] (5,-2) -- (5,2);
	\draw[dotted] (4,-2) -- (4,2);
        \draw[black,line width=1.5pt] (3,-2) -- (3,2);
        \draw[dotted] (2,-2) -- (2,2);
        \draw[dotted] (1,-2) -- (1,2);
        \draw[->] (0,-2)   -- (4,1);\etp \,\,\, +
\ldots\nonumber \\
 & =\sum_{k=1}^{N^{\rm I}}\left|\psi_{1}^{3}\ldots\phi_{k}^{a}\ldots\psi_{N^{\rm I}}^{3}\psi_{B}^{3}\psi_{N^{\rm I}}^{3}\ldots\psi_{1}^{3}\right\rangle \prod_{l=1}^{k-1}S^{\rm II,I}(y;x_{l})f^{in}(y;x_{k},\eta_{k}),\nonumber \\
 & \quad+\left|\psi_{1}^{3}\ldots\psi_{N^{\rm I}}^{3}{\phi_{B}^{a}}\psi_{N^{\rm I}}^{3}\ldots\psi_{1}^{3}\right\rangle \prod_{l=1}^{N^{\rm I}}S^{\rm II,I}(y;x_{l})f^{\tau}(y;x_{B},\eta_{B})\nonumber \\
 & \quad+\sum_{k=1}^{N^{\rm I}}\left|\psi_{1}^{3}\ldots\psi_{N^{\rm I}}^{3}\psi_{B}^{3}\psi_{N^{\rm I}}^{3}\ldots\phi_{k}^{a}\ldots\psi_{1}^{3}\right\rangle \prod_{l=1}^{N^{\rm I}}S^{\rm II,I}(y;x_{l})K^{\! \rm II}(y;x_{B})\nonumber \\
 & \qquad\qquad\qquad\qquad\qquad\qquad\qquad\qquad\!\times\prod_{l=k+1}^{N^{\rm I}}S^{\rm I,II}(x_{l};-y)f^{out}(x_{k},\eta_{k};-y)\label{single_L}.
\end{align}
Here  (see \cite{Beisert1} for the details)
\begin{equation*}
S^{\rm II,I}(y,x_{i}) =-\frac{y-x_{i}^{+}}{y-x_{i}^{-}},
\quad S^{\rm I,II}(x_{i};y)1/S^{\rm II,I}(y;x_{i}) =-\frac{y-x_{i}^{-}}{y-x_{i}^{+}},
\end{equation*}
\begin{equation}
f^{in}(y;x_{i},\eta_{i})=\frac{x_{i}^{+}-x_{i}^{-}}{y-x_{i}^{-}}\frac{1}{\eta_{i}},\quad
f^{out}(x_{i},\eta_{i};y)=S^{\rm I,II}(x_{i};y)\, f^{in}(y;x_{i},\eta_{i})=\frac{x_{i}^{-}-x_{i}^{+}}{y-x_{i}^{+}}\frac{1}{\eta_{i}}.
\label{levelIIS}\end{equation}
The new unknown functions $K^{\! \rm II}$ and $f^{\tau}$ are fixed
by the compatibility condition for the scattering through the boundary
\begin{align}
\mathcal{K}^{\rm I}\left|\Psi^{a}(y)\right\rangle ^{\! \rm II} & =\left|\Psi^{a}(y)\right\rangle _{\tau}^{\! \rm II}K^{\rm I}=\left|\Psi^{a}(y)\right\rangle _{\tau}^{\! \rm II}\nonumber \\
\mathcal{K}^{\rm I}\left|\Psi^{a}(y)\right\rangle ^{\! \rm II} & =\left|\Psi^{a}(y)\right\rangle _{\tau}^{\! \rm II}K^{\rm I}=\left|\Psi^{a}(y)\right\rangle _{\tau}^{\! \rm II},\label{compat_ref}
\end{align}
where $\tau$ merely acts by sending $x_{N^{\rm I}}^{\pm}\rightarrow-x_{N^{\rm I}}^{\mp}$.

We find\footnote{It is sufficient to consider a $N^{\rm I}=1$ state, i.e. one left, one right and one boundary site. A useful trick is to consider left-right graded-(anti)symmetric versions of the ansatz (\ref{single_L}):
\begin{align}
\left|\Psi^{\{a\}}(y)\right\rangle ^{\! \rm II} & =\mathbb{A}\Bigl(\left|\phi_{1}^{a}\psi_{B}^{3}\psi_{1}^{3}\right\rangle +\left|\phi_{1}^{3}\psi_{B}^{3}\psi_{1}^{a}\right\rangle \Bigr),\nonumber \\
\left|\Psi^{[a]}(y)\right\rangle ^{\! \rm II} & =\mathbb{B}\Bigl(\left|\phi_{1}^{a}\psi_{B}^{3}\psi_{1}^{3}\right\rangle -\left|\phi_{1}^{3}\psi_{B}^{3}\psi_{1}^{a}\right\rangle \Bigr)+\mathbb{C}\left|\phi_{1}^{3}\psi_{B}^{a}\psi_{1}^{3}\right\rangle ,
\end{align}
where
\begin{align}
\mathbb{A} & = \btp
	\draw[dotted] (3,-2) -- (3,2);
        \draw[black,line width=1.5pt] (2,-2) -- (2,2);
        \draw[dotted] (1,-2) -- (1,2);
         \draw[->] (0,-2)   -- (1,1);
\etp + \btp
	\draw[dotted] (3,-2) -- (3,2);
        \draw[black,line width=1.5pt] (2,-2) -- (2,2);
        \draw[dotted] (1,-2) -- (1,2);
         \draw[->] (0,-2)   -- (3,1);
\etp \,\,\, = \,\,\, \btp
	\draw[dotted] (3,-2) -- (3,2);
        \draw[black,line width=1.5pt] (2,-2) -- (2,2);
        \draw[dotted] (1,-2) -- (1,2);
         \draw[->] (4,-2)   -- (3,1);
\etp  \,\,\,+\,\,\, \btp
	\draw[dotted] (3,-2) -- (3,2);
        \draw[black,line width=1.5pt] (2,-2) -- (2,2);
        \draw[dotted] (1,-2) -- (1,2);
         \draw[->] (4,-2)   -- (1,1);
\etp \nonumber \\
 & = f^{in}(y;x_{1},\eta_{1})+S^{\rm II,I}(y;x_{1})K^{\rm II}(y;x_{B})f^{out}(x_{1},\eta_{1};-y),\nonumber \\
\mathbb{B} & = \btp
	\draw[dotted] (3,-2) -- (3,2);
        \draw[black,line width=1.5pt] (2,-2) -- (2,2);
        \draw[dotted] (1,-2) -- (1,2);
         \draw[->] (0,-2)   -- (1,1);
\etp \,\,\, - \btp
	\draw[dotted] (3,-2) -- (3,2);
        \draw[black,line width=1.5pt] (2,-2) -- (2,2);
        \draw[dotted] (1,-2) -- (1,2);
         \draw[->] (0,-2)   -- (3,1);
\etp \,\,\, =  \,\,\, \btp
	\draw[dotted] (3,-2) -- (3,2);
        \draw[black,line width=1.5pt] (2,-2) -- (2,2);
        \draw[dotted] (1,-2) -- (1,2);
         \draw[->] (4,-2)   -- (3,1);
\etp -\,\,\, \btp
	\draw[dotted] (3,-2) -- (3,2);
        \draw[black,line width=1.5pt] (2,-2) -- (2,2);
        \draw[dotted] (1,-2) -- (1,2);
         \draw[->] (4,-2)   -- (1,1);
\etp \nonumber \\
 & = f^{in}(y;x_{1},\eta_{1})-S^{\rm II,I}(y;x_{1})K^{\rm II}(y;x_{B})f^{out}(x_{1},\eta_{1};-y),\nonumber \\
\mathbb{C} & = \btp
	\draw[dotted] (3,-2) -- (3,2);
        \draw[black,line width=1.5pt] (2,-2) -- (2,2);
        \draw[dotted] (1,-2) -- (1,2);
         \draw[->] (0,-2)   -- (2,1);
\etp \,\,\,-\,\,\, \btp
	\draw[dotted] (3,-2) -- (3,2);
        \draw[black,line width=1.5pt] (2,-2) -- (2,2);
        \draw[dotted] (1,-2) -- (1,2);
         \draw[->] (4,-2)   -- (2,1);
\etp = 2\, S^{\rm II,I}(y;x_{1})f^{\tau}(y;x_{B},\eta_{B}).
\end{align}
The compatibility conditions (\ref{compat_ref}) then explicitly become
$k_{0}\,k_{7}^{\text{\tiny (S)}}\mathbb{A}=\mathbb{A}_{\tau}$,
for the graded-symmetric state and
$k_{3}^{\text{\tiny (A)}}\mathbb{B}+k_{18}^{\text{\tiny (A)}}\mathbb{C}=\mathbb{B}_{\tau}, 2k_{19}^{\text{\tiny (A)}}\mathbb{B}+k_{5}^{\text{\tiny (A)}}\mathbb{C}=\mathbb{C}_{\tau}$
for the graded-antisymmetric one.} the unique solution
\begin{equation}
K^{\rm II}(y;x_{B})=\frac{y-x_{B}}{y+x_{B}},\qquad f^{\tau}(y;x_{B},\eta_{B})=-\frac{\sqrt{2}x_{B}}{y+x_{B}}\frac{1}{\eta_{B}}. \label{KII}
\end{equation}

Note that we did not include terms $ \btp
	\draw[dotted] (4,-2) -- (4,2);
        \draw[black,line width=1.5pt] (3,-2) -- (3,2);
        \draw[dotted] (2,-2) -- (2,2);
        \draw[dotted] (1,-2) -- (1,2);
        \draw[-] (0,-2)   -- (3,0);
        \draw[->] (3,0)   -- (2,1); \etp \;$ and $\; \btp
	\draw[dotted] (4,-2) -- (4,2);
	\draw[dotted] (3,-2) -- (3,2);
        \draw[black,line width=1.5pt] (2,-2) -- (2,2);
        \draw[dotted] (1,-2) -- (1,2);
        \draw[-] (5,-2)   -- (2,0);
        \draw[->] (2,0)   -- (3,1); \etp $ in the ansatz. One could include such terms with some reflection
coefficient $\widetilde K^{\! \rm II}(y;x_{B})$, but solving the compatibility
relation (\ref{compat_ref}) one finds that $\widetilde K^{\! \rm II}(y;x_{B})=0$.
In this sense, the scattering from the boundary is indeed achiral: it is a sum of a left excitation with momentum $y$ and a right excitation with momentum $-y$, plus the boundary term.

For clarity in the pictures below, it is useful also to work with the spin-wave ansatz with its ``tail'' trailing to the right rather than the left -- that is, pictorially, $\ldots + \,\,\,\btp
	\draw[dotted] (5,-2) -- (5,2);
	\draw[dotted] (4,-2) -- (4,2);
        \draw[black,line width=1.5pt] (3,-2) -- (3,2);
        \draw[dotted] (2,-2) -- (2,2);
        \draw[dotted] (1,-2) -- (1,2);
        \draw[->] (6,-2)   -- (4,1);\etp
+ \,\,\, \btp
	\draw[dotted] (5,-2) -- (5,2);
	\draw[dotted] (4,-2) -- (4,2);
        \draw[black,line width=1.5pt] (3,-2) -- (3,2);
        \draw[dotted] (2,-2) -- (2,2);
        \draw[dotted] (1,-2) -- (1,2);
        \draw[->] (6,-2)   -- (3,1);\etp
+ \,\,\, \btp
	\draw[dotted] (5,-2) -- (5,2);
	\draw[dotted] (4,-2) -- (4,2);
        \draw[black,line width=1.5pt] (3,-2) -- (3,2);
        \draw[dotted] (2,-2) -- (2,2);
        \draw[dotted] (1,-2) -- (1,2);
        \draw[->] (6,-2)   -- (2,1);\etp + \ldots$.
However, such states are not linearly independent of those of the form (\ref{single_L}). Thus, in contrast to the usual open boundaries case, there is only one type of level II excitation.

\paragraph{Scattering of level II particles.} The scattering of two \textit{level II} excitations in the bulk works as in the usual open-boundaries case; neglecting the boundary,  the \textit{level II} state of two particles is in a background consisting of two \textit{level I} sites is \begin{align}
\left|\phi^{a}(y_{1})\phi^{b}(y_{2})\right\rangle _{\text{\scriptsize two site}}^{\! \rm II} & = \mathbb{A}\left|\phi_{1}^{a}\phi_{2}^{b}\right\rangle \nonumber + \mathbb{B}\left(M\left|\phi_{1}^{a}\phi_{2}^{b}\right\rangle +N\left|\phi_{1}^{b}\phi_{2}^{a}\right\rangle \right)\nonumber \\
 & \qquad\qquad\qquad+\varepsilon^{ab}\left(\mathbb{C}\left|\psi_{1}^{4}\psi_{2}^{3}\right\rangle +\mathbb{D}\left|\psi_{1}^{3}\psi_{2}^{4}\right\rangle \right),\end{align}
where the shorthands are
\begin{align}
\mathbb{A} & = \btp
	\draw[dotted] (3,-2) -- (3,2);
        \draw[dotted] (2,-2) -- (2,2);
        \draw[->] (1,-2)   -- (3,1);
	\draw[->] (0,-2)   -- (2,1);
\etp \,\,\, = \; f^{in}(y_{1};x_{1},\eta_{1})S^{\rm II,I}(y_{2};x_{1})f^{in}(y_{2};x_{2},\eta_{2}),\nonumber \\
\mathbb{B} & = \btp
	\draw[dotted] (3,-2) -- (3,2);
        \draw[dotted] (2,-2) -- (2,2);
        \draw[->] (1,-2)   -- (2,1);
	\draw[->] (0,-2)   -- (3,1);
\etp \,\,\, = \; S^{\rm II,I}(y_{1};x_{1}) f^{in}(y_{1};x_{2},\eta_{2}) f^{in}(y_{2};x_{1},\eta_{1}),\nonumber \\
\mathbb{C} & = \btp
	\draw[dotted] (3,-2) -- (3,2);
        \draw[dotted] (2,-2) -- (2,2);
	\filldraw[fill=black] (2,1) circle (1.5mm);
        \draw[-] (1,-2)   -- (2,1);
	\draw[-] (0,-2)   -- (2,1);
\etp \,\,\, = \; f^{in}(y_{1};x_{1},\eta_{1}) f^{in}(y_{2};x_{1},\eta_{1}) f^{\sigma}(y_{1},y_{2};x_{1},\eta_{1},\zeta_{1}),\nonumber \\
\mathbb{D} & = \btp
	\draw[dotted] (3,-2) -- (3,2);
        \draw[dotted] (2,-2) -- (2,2);
	\filldraw[fill=black] (3,1) circle (1.5mm);
        \draw[-] (1,-2)   -- (3,1);
	\draw[-] (0,-2)   -- (3,1);
\etp \,\,\, = \; S^{\rm II,I}(y_{1};x_{1})f^{in}(y_{1};x_{1},\eta_{1})S^{\rm II,I}(y_{2};x_{1})f^{in}(y_{2};x_{1},\eta_{1})f^{\sigma}(y_{1},y_{2};x_{2},\eta_{2},\zeta_{2}).
\end{align}
Here \cite{Beisert1}
\begin{equation}
f^{\sigma}(y_{1},y_{2};x_{k},\eta_{k},\zeta_{k})=
  \frac{x_{k}^{+}x_{k}^{-}-y_{1}y_{2}}{x_{k}^{+}(x_{k}^{+}-x_{k}^{-})}\frac{\eta_{k}}{\zeta_{k}}\left(-\frac{\frac{i}{y_{1}}-\frac{i}{y_{2}}}{v_{1}-v_{2}-\frac{2i}{g}}\right),
\end{equation}
and
\begin{equation}
M(y_{1},y_{2})=\frac{\frac{2i}{g}}{v_{1}-v_{2}-\frac{2i}{g}},\quad N(y_{1},y_{2})=-\frac{v_{1}-v_{2}}{v_{1}-v_{2}-\frac{2i}{g}},\label{MNfS}
\end{equation}
where $v_{i}=y_{i}+1/y_{i}$.

In the presence of a boundary there is only one new type of term needed in the ansatz, corresponding to both particles sitting at the boundary. The full two-particle ansatz has many terms, so for brevity we shall write it out only in the case of a chain with $N^{\rm I}=1$ bulk sites. This is sufficient to determine the new coefficient, $f^{\sigma\tau}$.

Thus, consider a state of two impurities propagating
on a background of a three-site \textit{level I} chain of left, right and
boundary slots. For clarity in the pictures we suppose that the impurities have their ``tails'' running one to the left and one to the right.
The \textit{level II} ansatz is
\begin{align}
\left|\phi^{a}(y_{1})\phi^{b}(y_{2})\right\rangle _{\text{\scriptsize three site}}^{\! \rm II}
 & =\mathbb{A}\left|\phi_{1}^{a}\psi_{B}^{3}\phi_{2}^{b}\right\rangle +\mathbb{B}\left|\phi_{1}^{a}\psi_{B}^{b}\phi_{2}^{3}\right\rangle -\mathbb{C}\left|\phi_{1}^{3}\psi_{B}^{a}\phi_{2}^{b}\right\rangle \nonumber \\
 & \quad+M(y_{1},-y_{2})\left(\mathbb{D}\left|\phi_{1}^{a}\psi_{B}^{3}\phi_{2}^{b}\right\rangle +\mathbb{E}\left|\phi_{1}^{a}\psi_{B}^{b}\phi_{2}^{3}\right\rangle -\mathbb{F}\left|\phi_{1}^{3}\psi_{B}^{a}\phi_{2}^{b}\right\rangle \right)\nonumber \\
 & \quad+N(y_{1},-y_{2})\left(\mathbb{D}\left|\phi_{1}^{b}\psi_{B}^{3}\phi_{2}^{a}\right\rangle +\mathbb{E}\left|\phi_{1}^{b}\psi_{B}^{a}\phi_{2}^{3}\right\rangle -\mathbb{F}\left|\phi_{1}^{3}\psi_{B}^{a}\phi_{2}^{b}\right\rangle \right)\nonumber \\
 & \quad+\varepsilon^{ab}\left(\mathbb{G}\left|\psi_{1}^{4}\psi_{B}^{3}\psi_{2}^{3}\right\rangle +\mathbb{H}\left|\psi_{1}^{3}\psi_{B}^{4}\psi_{2}^{3}\right\rangle +\mathbb{K}\left|\psi_{1}^{3}\psi_{B}^{3}\psi_{2}^{4}\right\rangle \right),\label{ab_bd}
\end{align}
where the shorthands are
\begin{align}
\mathbb{A} & = \btp
	\draw[dotted] (3,-2) -- (3,2);
        \draw[black,line width=1.5pt] (2,-2) -- (2,2);
        \draw[dotted] (1,-2) -- (1,2);
        \draw[->] (0,-2)   -- (1,1);
	\draw[->] (4,-2)   -- (3,1);
\etp = f^{in}(y_{1};x_{1},\eta_{1})f^{in}(y_{2};x_{2},\eta_{2}),\nonumber \\
\mathbb{B} & = \btp
	\draw[dotted] (3,-2) -- (3,2);
        \draw[black,line width=1.5pt] (2,-2) -- (2,2);
        \draw[dotted] (1,-2) -- (1,2);
        \draw[->] (0,-2)   -- (1,1);
	\draw[->] (4,-2)   -- (2,1);
\etp = -f^{in}(y_{1};x_{1},\eta_{1})S^{\rm II,I}(y_{2};x_{2})f^{\tau}(y_{2};x_{B},\eta_{B}),\nonumber \\
\mathbb{C} & = \btp
	\draw[dotted] (3,-2) -- (3,2);
        \draw[black,line width=1.5pt] (2,-2) -- (2,2);
        \draw[dotted] (1,-2) -- (1,2);
        \draw[->] (0,-2)   -- (2,1);
	\draw[->] (4,-2)   -- (3,1);
\etp = S^{\rm II,I}(y_{1};x_{1})f^{\tau}(y_{1};x_{B},\eta_{B})f^{in}(y_{2};x_{2},\eta_{2}),\nonumber \\
\mathbb{D} & = \btp
	\draw[dotted] (3,-2) -- (3,2);
        \draw[black,line width=1.5pt] (2,-2) -- (2,2);
        \draw[dotted] (1,-2) -- (1,2);
        \draw[->] (0,-2)   -- (3,1);
	\draw[->] (4,-2)   -- (1,0);
\etp = S^{\rm II,I}(y_{1};x_{1})K^{\! \rm II}(y_{1},x_{B})f^{out}(x_{2},\eta_{2};-y_{1}) \nonumber \\
 & \qquad\qquad\qquad \times S^{\rm II,I}(y_{2};x_{2})K^{\rm II,I}(y_{2},x_{B})f^{out}(x_{1},\eta_{1};-y_{2}),\nonumber \\
\mathbb{E} & = \btp
	\draw[dotted] (3,-2) -- (3,2);
        \draw[black,line width=1.5pt] (2,-2) -- (2,2);
        \draw[dotted] (1,-2) -- (1,2);
        \draw[->] (0,-2)   -- (2,1);
	\draw[->] (4,-2)   -- (1,0.5);
\etp = S^{\rm II,I}(y_{1};x_{1})f^{\tau}(y_{1};x_{B},\eta_{B})S^{\rm II,I}(y_{2};x_{2})K^{\rm II,I}(y_{2},x_{B})f^{out}(x_{1},\eta_{1};-y_{2}),\nonumber \\
\mathbb{F} & = \btp
	\draw[dotted] (3,-2) -- (3,2);
        \draw[black,line width=1.5pt] (2,-2) -- (2,2);
        \draw[dotted] (1,-2) -- (1,2);
        \draw[->] (0,-2)   -- (3,0.5);
	\draw[->] (4,-2)   -- (2,1);
\etp = S^{\rm II,I}(y_{1};x_{1})K^{\rm II,I}(y_{1},x_{B})f^{out}(x_{2},\eta_{2};-y_{1})S^{\rm II,I}(y_{2};x_{2})f^{\tau}(y_{2};x_{B},\eta_{B}),\nonumber \\
\mathbb{G} & = \btp
	\draw[dotted] (3,-2) -- (3,2);
        \draw[black,line width=1.5pt] (2,-2) -- (2,2);
        \draw[dotted] (1,-2) -- (1,2);
	\filldraw[fill=black] (1,1) circle (1.5mm);
        \draw[-] (0,-2)   -- (1,1);
 	\draw[-] (4,-2)   -- (1,1);
\etp = f^{in}(y_{1};x_{1},\eta_{1})S^{\rm II,I}(y_{2};x_{2})K^{\rm II,I}(y_{2},x_{B})f^{out}(x_{1},\eta_{1};-y_{2})f^{\sigma}(y_{1},-y_{2};x_{1},\eta_{1},\zeta_{1}),\nonumber \\
\mathbb{H} & = \btp
	\draw[dotted] (3,-2) -- (3,2);
        \draw[black,line width=1.5pt] (2,-2) -- (2,2);
        \draw[dotted] (1,-2) -- (1,2);
	\filldraw[fill=black] (2,1) circle (1.5mm);
        \draw[-] (0,-2)   -- (2,1);
 	\draw[-] (4,-2)   -- (2,1);
\etp = S^{\rm II,I}(y_{1};x_{1})f^{\tau}(y_{1};x_{B},\eta_{B})S^{\rm II,I}(y_{2};x_{2})f^{\tau}(y_{2};x_{B},\eta_{B})f^{\sigma\tau}(y_{1},y_{2};x_{B},\eta_{B},\zeta_{B}),\nonumber \\
\mathbb{K} & = \btp
	\draw[dotted] (3,-2) -- (3,2);
        \draw[black,line width=1.5pt] (2,-2) -- (2,2);
        \draw[dotted] (1,-2) -- (1,2);
	\filldraw[fill=black] (3,1) circle (1.5mm);
        \draw[-] (0,-2)   -- (3,1);
 	\draw[-] (4,-2)   -- (3,1);
\etp = S^{\rm II,I}(y_{1};x_{1})K^{\rm II,I}(y_{1},x_{B})f^{out}(x_{2},\eta_{2};-y_{1})f^{in}(y_{2};x_{1},\eta_{1})f^{\sigma}(-y_{1},y_{2};x_{2},\eta_{2},\zeta_{2}).\end{align}
The minus signs in (\ref{ab_bd}) appear because of the graded permutation of left and right representations, which is not explicitly seen in the unfolded picture, but is revealed by folding the incoming left tail to the right side:
\begin{equation*}
- \btp
	\draw[dotted] (5,-2) -- (5,2);
	\draw[dotted] (4,-2) -- (4,2);
        \draw[black,line width=1.5pt] (3,-2) -- (3,2);
        \draw[dotted] (2,-2) -- (2,2);
        \draw[dotted] (1,-2) -- (1,2);
        \draw[->] (0,-2)   -- (4,1);
	\draw[->] (6,-2)   -- (3,1);
\etp \sim \;\; \btp
	\draw[dotted] (5,-2) -- (5,2);
	\draw[dotted] (4,-2) -- (4,2);
        \draw[black,line width=1.5pt] (3,-2) -- (3,2);
        \draw[->] (3,0)   -- (4,1);
        \draw[-]  (3,0)   -- (7,-2);
	\draw[->] (6,-2)   -- (3,1);
\etp, \qquad \text{and} \qquad - \btp
	\draw[dotted] (5,-2) -- (5,2);
	\draw[dotted] (4,-2) -- (4,2);
        \draw[black,line width=1.5pt] (3,-2) -- (3,2);
        \draw[dotted] (2,-2) -- (2,2);
        \draw[dotted] (1,-2) -- (1,2);
	\filldraw[fill=black] (4,1) circle (1.5mm);
        \draw[-] (0,-2)   -- (4,1);
	\draw[-] (6,-2)   -- (4,1);
\etp \sim \;\; \btp
	\draw[dotted] (5,-2) -- (5,2);
	\draw[dotted] (4,-2) -- (4,2);
        \draw[black,line width=1.5pt] (3,-2) -- (3,2);
	\filldraw[fill=black] (4,1) circle (1.5mm);
        \draw[-] (3,0)   -- (4,1);
        \draw[-] (3,0)   -- (7,-2);
	\draw[-] (6,-2)   -- (4,1);
\etp.
\end{equation*}
The minus sign in $\mathbb B$ is already present in the ansatz for a single \textit{level II} particle with its tail running to the right.

The compatibility relation for the case under the consideration is\begin{align}
\mathcal{K}^{\rm I}\left|\phi^{a}(y_{1})\phi^{b}(y_{2})\right\rangle ^{\! \rm II} & =\left|\phi^{a}(y_{1})\phi^{b}(y_{1})\right\rangle _{\tau}^{\! \rm II}K^{\rm I}=\left|\phi^{a}(y_{1})\phi^{b}(y_{2})\right\rangle _{\tau}^{\! \rm II},\end{align}
and is trivially satisfied when $a=b$, because the terms with
unknown function $f^{\sigma\tau}$ do not appear. It is good to
start from this case, as it is a way to check that the ansatz is written
consistently and all signs are correct.

The easiest way to find $f^{\sigma\tau}$ it is to consider
the overlap of the consistency condition with the composite excitation
$\left|\psi^{4}(y_{1},y_{2})\right\rangle _{\text{\scriptsize three site}}^{\! \rm II}$
leading to a very simple equation
\begin{equation}
k_{5}^{\text{\tiny (A)}}(\mathbb{G}+\mathbb{H}-\mathbb{K})=\mathbb{G}_{\tau}+\mathbb{H}_{\tau}-\mathbb{K}_{\tau},
\end{equation}
which may be solved straightforwardly giving a unique solution
\begin{equation}
f^{\sigma\tau}(y_{1},y_{2},x_{B},\eta_{B},\zeta_{B})=-i\frac{\eta_{B}^{2}}{\zeta_{B}}\frac{\Bigl(\frac{1}{y_{1}}+\frac{1}{y_{2}}\Bigr)\Bigl(1-\frac{y_{1}y_{2}}{x_{B}^{2}}\Bigr)}{v_{1}+v_{2}-\frac{2i}{g}}.
\end{equation}
Then it is easy to check that all other constrains coming from the
compatibility relation are satisfied using this result and the mass-shell
relation.

Note that we did not include the following diagram in (\ref{ab_bd}). $ \btp
	\draw[dotted] (5,-2) -- (5,2);
	\draw[dotted] (4,-2) -- (4,2);
        \draw[black,line width=1.5pt] (3,-2) -- (3,2);
        \draw[dotted] (2,-2) -- (2,2);
        \draw[dotted] (1,-2) -- (1,2);
        \draw[->] (0,-2)   -- (4,0.3);
	\draw[->] (6,-2)   -- (2,1); \etp $.
It is a valid scattering diagram, but it does not need to be included
as it is equivalent to the diagram we have already included, namely $ \btp
	\draw[dotted] (5,-2) -- (5,2);
	\draw[dotted] (4,-2) -- (4,2);
        \draw[black,line width=1.5pt] (3,-2) -- (3,2);
        \draw[dotted] (2,-2) -- (2,2);
        \draw[dotted] (1,-2) -- (1,2);
        \draw[->] (0,-2)   -- (4,1);
	\draw[->] (6,-2)   -- (2,0.3); \etp \;$. (The
\textit{level II} reflection matrix has only one component, namely $K^{\rm II}$ (\ref{KII}), and
thus acts only diagonally.)

\subsubsection*{Level III}
The final level of the nesting is very similar to that in \cite{Correa2}. One finds the usual \textit{level III} S-matrices of \cite{Beisert1}:
\begin{equation}
S^{\rm III,II}(w;y)=\frac{w-v+\frac{i}{g}}{w-v-\frac{i}{g}},\quad
S^{\rm III}(w_{1},w_{2})=\frac{w_{1}-w_{2}-\frac{2i}{g}}{w_{1}-w_{2}+\frac{2i}{g}}.
\label{SIII}
\end{equation}
and the \textit{level III} reflection matrix
\begin{equation}
K^{\rm III}(w)=-1.
\label{KIII}
\end{equation}

\subsection{Bethe equations}

\begin{figure}[ht]
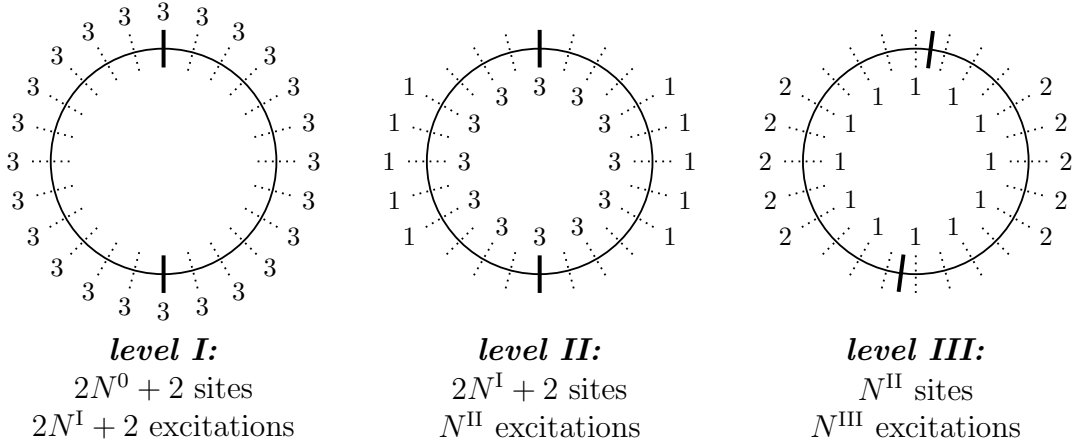
 \begin{center} \btp

  \draw (0,7) circle (6);
  \foreach \angle in {-75, -60, ..., 75, 105, 120, ..., 255}
    \draw [dotted,rotate around={\angle:(0,7)}] (5,7) -- (7,7);
  \foreach \angle in {-75, -60, ..., 75}
    \draw [rotate around={\angle:(0,7)}] (8,7) node {$\text{\footnotesize{3}}$};
  \foreach \angle in {105, 120, ..., 255}
    \draw [rotate around={\angle:(0,7)}] (8,7) node {$\text{\footnotesize{3}}$};
  \foreach \angle in {90,270}
    \draw [rotate around={\angle:(0,7)}] (8,7) node {$\text{\footnotesize{3}}$};
  \draw[black,line width=1.5pt] (0,12) -- (0,14);
  \draw[black,line width=1.5pt] (0,2) -- (0,0);

  \draw (20,7) circle (6);
  \foreach \angle in {-75, -60, ..., 75, 105, 120, ..., 255}
     \draw [dotted,rotate around={\angle:(20,7)}] (25,7) -- (27,7);
  \foreach \angle in {-60, -30, ..., 60}
     \draw [dotted,rotate around={\angle:(20,7)}] (24,7) node {$\text{\footnotesize{3}}$};
  \foreach \angle in {120, 150, ..., 240}
     \draw [dotted,rotate around={\angle:(20,7)}] (24,7) node {$\text{\footnotesize{3}}$};
  \foreach \angle in {90,270}
     \draw [dotted,rotate around={\angle:(20,7)}] (24,7) node {$\text{\footnotesize{3}}$};
  \foreach \angle in {210,195,...,150}
    \draw [dotted,rotate around={\angle:(20,7)}] (28,7) node {$\text{\footnotesize{1}}$};
  \foreach \angle in {-30,-15,...,30}
    \draw [dotted,rotate around={\angle:(20,7)}] (28,7) node {$\text{\footnotesize{1}}$};
  \draw[black,line width=1.5pt] (20,12) -- (20,14);
  \draw[black,line width=1.5pt] (20,2) -- (20,0);

  \draw (40,7) circle (6);
  \foreach \angle in {0, 15, ..., 345}
    \draw [dotted,rotate around={\angle:(40,7)}] (45,7) -- (47,7);
  \foreach \angle in {-90, -60, ..., 60}
    \draw [dotted,rotate around={\angle:(40,7)}] (44,7) node {$\text{\footnotesize{1}}$};
  \foreach \angle in {90, 120, ..., 240}
    \draw [dotted,rotate around={\angle:(40,7)}] (44,7) node {$\text{\footnotesize{1}}$};
  \foreach \angle in {210,195,...,150}
    \draw [dotted,rotate around={\angle:(40,7)}] (48,7) node {$\text{\footnotesize{2}}$};
  \foreach \angle in {-30,-15,...,30}
    \draw [dotted,rotate around={\angle:(40,7)}] (48,7) node {$\text{\footnotesize{2}}$};
  \draw[black,line width=1.5pt,rotate around={-7.5:(40,7)}] (40,12) -- (40,14);
  \draw[black,line width=1.5pt,rotate around={-7.5:(40,7)}] (40,2)  -- (40,0);

  \draw (0,-3) node {\textbf{\textit{level I:}}} (0,-5) node {$2N^{0}+2$ sites} (0,-7) node {$2N^{\rm I}+2$ excitations}
	(20,-3) node {\textbf{\textit{level II:}}} (20,-5) node {$2N^{\rm I}+2$ sites} (20,-7) node {$N^{\rm II}$ excitations}
	(40,-3) node {\textbf{\textit{level III:}}} (40,-5) node {$N^{\rm II}$ sites} (40,-7) node {$N^{\rm III}$ excitations};

\etp \end{center}
\caption{ Schematical representation of Bethe equations for all levels. The dotted lines represent bulk sites and the solid lines represent boundary sites. The inside of the circles corresponds to the background states and the outside corresponds to the excitations.} \label{BEs}
\end{figure}

\noindent The nested coordinate Bethe ansatz  we have presented applies to a semi-infinite
system with a right boundary. The picture was unfolded into an infinite system with the right boundary represented in the middle. Adding a left boundary at a distance $L$, corresponds to closing the infinite line into a circle of length $2L$. Then the Bethe equations are obtained by inserting excitations into the closed spin-chain at any level (I, II or III) and moving them around the circle, scattering them with all other states and with left and right boundaries (see figure \ref{BEs}).

The \textit{level I} chain has $2 N^{0} + 2$ sites on which to place  $N^{\rm I}$ left,
$N^{\rm I}$ right, and two boundary excitations. The full revolution of any
\textit{level I} excitation around the circle results in a phase factor $e^{2ipL}$, where $L$ is
`a half of the circumference of the circle'. 

At \textit{level II} there are $N^{\rm II}$ excitations that propagate in the
inhomogeneous background of the $2N^{\rm I}+2$ \textit{level I} sites. Thus there can be any number  $N^{\rm II}\in [0,2N^{\rm I}+2]$ of \textit{level II} states.
The final level, \textit{level III}, is similar, but there are no boundary sites. There can be any number $N^{\rm III}\in[0,N^{\rm II}]$ of \textit{level III} excitations.

Each excitation of every level has  a rapidity $x_{k}^{A}$ associated, and for each one of them we obtain a Bethe equation:
\begin{equation}\def\arraystretch{1}
K_{R}^{A}(x_{k}^{A})\,K_{L}^{A}(-x_{k}^{A}) \!\!\underset{(A,k)\neq (B,\ell)}{\prod_{B=I}^{III}\prod_{l=1}^{N^{B}}}\!\!
    S^{A,B}(x_{k}^{A},x_{l}^{B})\, S^{B,A}(x_{l}^{B},-x_{k}^{A})
= \left\{\begin{array}{cl} \left(\frac{x^{+}_k}{x^{-}_k}\right)^{-2L} & {\rm for\ } A={\rm I} \\
1 & {\rm for\ } A=\rm{II, III}.
\end{array}\right.
\end{equation}

For \textit{level I} rapidities we use the  spectral parameters $x^{\rm I} = x^\pm$ and for \textit{level II} and \textit{level III} rapidities we use $y$ and $w$ respectively.
We can now write the Bethe equations explicitly for each level using
(\ref{levelI}), (\ref{levelIIS}), (\ref{KII}), (\ref{SIII}) and (\ref{KIII}) and simplify them
with the help of parity symmetry, which  ensures that $K^A_L(-x^A) = K^A_R(x^A)$
and $S^{B,A}(x_{l}^{B},-x_{k}^{A}) = S^{A,B}(x_{k}^{A},-x_{l}^{B})$.
Then, the expressions of the Bethe equations for the $\mathfrak{su}(2|2)^2$
scattering theory with the ``achiral vertical'' boundary conditions are
\begin{align}
1 &= K_{0}(x_{k}^{\pm})^{2}\left(\frac{x_{k}^{+}}{x_{k}^{-}}\right)^{2L}\prod_{l\neq k}^{N^{{\rm I}}}S_{0}(x_{k}^{\pm},x_{l}^{\pm})^{2}S_{0}(x_{k}^{\pm},-x_{l}^{\mp})^{2}
   \prod_{l=1}^{N^{\rm II}}\frac{y_{l}-x_{k}^{-}}{y_{l}-x_{k}^{+}}\frac{y_{l}+x_{k}^{-}}{y_{l}+x_{k}^{+}}, \label{BI}\\
1 &= \left(\frac{y_{k}-x_{B}}{y_{k}+x_{B}}\right)^{2}\prod_{l=1}^{N^{{\rm I}}}\frac{y_{k}-x_{l}^{+}}{y_{k}-x_{l}^{-}}\frac{y_{k}+x_{l}^{-}}{y_{k}+x_{l}^{+}}
    \prod_{l=1}^{N^{\rm III}}\frac{w_{l}-v_{k}-\frac{i}{g}}{w_{l}-v_{k}+\frac{i}{g}}\frac{w_{l}+v_{k}+\frac{i}{g}}{w_{l}+v_{k}-\frac{i}{g}}, \label{BII}\\
1 &= \prod_{l=1}^{N^{\rm II}}\frac{w_{k}-v_{l}+\frac{i}{g}}{w_{k}-v_{l}-\frac{i}{g}}\frac{w_{k}+v_{l}-\frac{i}{g}}{w_{k}+v_{l}+\frac{i}{g}}
    \prod_{l\neq k}^{N^{\rm III}}\frac{w_{k}-v_{l}-\frac{2i}{g}}{w_{k}-w_{l}+\frac{2i}{g}}\frac{w_{k}+w_{l}-\frac{2i}{g}}{w_{k}+w_{l}+\frac{2i}{g}} \label{BIII}.
\end{align}

Note that we added the overall scalar factors $K_{0}(x_{k}^{\pm})$ and $S_{0}(x_{k}^{\pm},x_{l}^{\pm})^{2}$ to the \textit{level I} scattering factors (\ref{levelI}), which are not determined by symmetry arguments. The bulk factor $S_{0}(x_{k}^{\pm},x_{l}^{\pm})^{2}$ was found by an educated guess which relied on the crossing symmetry and many sophisticated weak and strong coupling verifications \cite{Janik,BHL,BESc}. However, the analogous boundary factor $K_{0}(x_{k}^{\pm})$ for the $D5$-brane reflection is not known yet.

Note the similarity between the Bethe equations (\ref{BI})-(\ref{BIII})
and the Bethe equations for the $\mathfrak{su}(2|2)$ scattering theory with a ``$Z=0$'' boundary \cite{Correa2}. The only differences are in the dressing phase $K_{0}$, and that the equations lack an index $\alpha$ in \textit{level II} and \textit{level III} rapidities, which distinguishes between left and right excitations. This is because the achiral nature of the reflection, which means that left and right can no longer be distinguished.

The flavour of \textit{level I} excitations is a matter of choice. If we had we chosen them to be $\phi^1$ instead of $\psi^3$, the $\psi^\alpha$ would have been the \textit{level II} excitations and for the scattering factors of the nested Bethe ansatz we would have obtained: \begin{align}
& S^{\rm I}(x_{1};x_{2})=\frac{x_{1}^{-}-x_{2}^{+}}{x_{1}^{+}-x_{2}^{-}}\sqrt{\frac{x_{1}^{+}x_{2}^{-}}{x_{1}^{-}x_{2}^{+}}},
 && K^{\rm I}(x;x_{B})=-\frac{x^{+}}{x^{-}}\left(\frac{x^{-}-x_{B}}{x^{+}+x_{B}}\right)^{2},\nonumber \\
&  K^{\rm II}(y;x_{B})=\frac{y+x_B}{y-x_B},
 && S^{\rm II,I}(y;x)=\frac{y-x^{-}}{y-x^{+}},\nonumber \\
& S^{\rm III,II}(w;v)=\frac{w-v-\frac{i}{g}}{w-v+\frac{i}{g}},
 && S^{\rm III}(w_{1};w_{2})=\frac{w_{1}-w_{2}+\frac{2i}{g}}{w_{1}-w_{2}-\frac{2i}{g}}.
\end{align}
Thus the Bethe equations would be of the following form:
\begin{align}
1 &= K_{0}(x_{k}^{\pm})^{2}\left(\frac{x_{k}^{+}}{x_{k}^{-}}\right)^{2L}\left(\frac{x_{k}^{+}}{x_{k}^{-}}\right)^{2}\left(\frac{x_{k}^{-}-x_{B}}{x_{k}^{+}+x_{B}}\right)^{4}\nonumber\\
  &  \qquad\times\prod_{l\neq k}^{N^{{\rm I}}}S_{0}(x_{k}^{\pm},x_{l}^{\pm})^{2}S_{0}(x_{k}^{\pm},-x_{l}^{\mp})^{2}\left(\frac{x_{k}^{-}-x_{l}^{+}}{x_{k}^{+}-x_{l}^{-}}\sqrt{\frac{x_{k}^{+}x_{l}^{-}}{x_{k}^{-}x_{l}^{+}}}\right)^{2}\left(\frac{x_{k}^{-}+x_{l}^{-}}{x_{k}^{+}+x_{l}^{+}}\sqrt{\frac{x_{k}^{+}x_{l}^{+}}{x_{k}^{-}x_{l}^{-}}}\right)^{2}\nonumber\\
  &  \qquad\times\prod_{l=1}^{N^{{\rm II}}}\frac{y_{l}-x_{k}^{+}}{y_{l}-x_{k}^{-}}\frac{y_{l}+x_{k}^{+}}{y_{l}+x_{k}^{-}},\label{BI_vac1} \\
1 &= \left(\frac{y_{k}+x_{B}}{y_{k}-x_{B}}\right)^{2}\prod_{l=1}^{N^{{\rm I}}}\frac{y_{k}-x_{l}^{-}}{y_{k}-x_{l}^{+}}\frac{y_{k}+x_{l}^{+}}{y_{k}+x_{l}^{-}}\prod_{l=1}^{N^{{\rm III}}}\frac{w_{l}-v_{k}+\frac{i}{g}}{w_{l}-v_{k}-\frac{i}{g}}\frac{w_{l}+v_{k}-\frac{i}{g}}{w_{l}+v_{k}+\frac{i}{g}},\label{BII_vac1}\\
1 &= \prod_{l=1}^{N^{{\rm II}}}\frac{w_{k}-v_{l}-\frac{i}{g}}{w_{k}-v_{l}+\frac{i}{g}}\frac{w_{k}+v_{l}+\frac{i}{g}}{w_{k}+v_{l}-\frac{i}{g}}\prod_{l\neq k}^{N^{{\rm III}}}\frac{w_{k}-v_{l}+\frac{2i}{g}}{w_{k}-w_{l}-\frac{2i}{g}}\frac{w_{k}+w_{l}+\frac{2i}{g}}{w_{k}+w_{l}-\frac{2i}{g}}.\label{BIII_vac1}
\end{align}

As a matter of convention, we take the distance $L$ to be the number of bulk fields $N^0$. After having made this choice, the overall factor $K_{0}$ could be fixed order by order by comparison with weak and strong coupling results.

Given that the spin-chain length may vary under mixing \cite{Beisert1}, $N^0$ is not
a good quantum number. As  in \cite{Correa2}, the Bethe equations
can be re-expressed in terms of the charge $J=J_{\rm 56}$. The expression of the conserved charged
in terms of the $N^A$ depends on the vacuum orientation. For the vacuum $\psi^3$
choice this would be $J \sim N^0- N^{\rm II}$ while for the vacuum $\phi^1$ would be
$J \sim N^0-N^{\rm I}+ N^{\rm II}$.

\paragraph{Horizontal vacuum.} Let us go back and consider
the reflection in the horizontal vacuum case. We had shown that this case is equivalent to the scattering in the bulk (as in  \cite{Beisert1}).
The Bethe equations for the boundary scattering in the horizontal case
are essentially the same as in the vertical case. For instance, eq. (\ref{BI})-(\ref{BIII}) would
remain the same, except for the factor $\tfrac{y_k-x_B}{y_k+x_B}$ in (\ref{BII}) which would not appear. Of course, the boundary dressing phase would also be different. The analogue of eq. (\ref{BI})-(\ref{BIII}) in the horizontal case would be identical to the Bethe equations of a closed spin chain of length $2L$ with $2N^{\rm I}$ bulk magnons of momenta $(p_1,\dots,p_{N^{\rm I}},-p_{N^{\rm I}},\dots,-p_1)$ \emph{if} the boundary dressing factor in this case was $K_0(p)=S_0(p,-p)$, but that has not been proven.


\section{Discussion}
\label{discussion}

In this paper we have considered the reflection of fundamental magnons from certain $D5$-branes in an $AdS_5\times S^5$ background.
There are two interesting cases of embedding $D5$-branes that have different vacuum orientations, which we have named ``horizontal'' and ``vertical'' vacua. A previous attempt to show the integrability of these configurations \cite{Correa1} was unsuccessful because some crucial minus signs were overlooked. In this paper we have shown that the $D5$-brane, from the scattering theory point of view, allows integrable boundary conditions and furthermore, by solving Bethe ansatz equations, we have shown that they are of the achiral (chirality-reversing) type.

Interestingly, the reflection from the achiral boundary has an ``unfolded'' picture where the reflection from the boundary may be considered as a ``scattering through the boundary''. In this ``unfolded'' picture, the scattering in the ``horizontal'' vacuum open spin-chain is completely equivalent to the scattering in an closed spin-chain. This is so because the boundary is achiral and has no boundary degrees of freedom (is a singlet). Thus, the Bethe equations are essentially the same, up to a possibly different dressing factor.

Some interesting calculations could be done to verify the assertion of all-loop integrability for the considered $D$5-brane boundary conditions. In the weak coupling limit, one could perform a 2-loop computation to analyze whether the dilatation operator is integrable beyond 1-loop. In the opposite regime, in the strong coupling, one should expect the superstring theory to be classically integrable. In particular, it would be very interesting to understand why the infinite set of non-local charges could be constructed for the $D$3 and the $D$7 case but failed for the $D$5 in the formalism of \cite{MaVa}. As a separate question, one could also compute the leading finite size corrections, as was done for the $D$3 case in \cite{CY}.

Although we have analyzed vertical and horizontal cases separately, the integrability of both cases ought to be related. Consider the open spin chain corresponding to our $D$5-branes. If we knew its exact anomalous dimension Hamiltonian, integrability would depend on the existence of conserved charges only, independently of any vacuum choice. However, while the reflection matrix in the horizontal case necessarily satisfied the bYBE, in the vertical case the boundary symmetry constraints were not enough to fix the reflection matrix to satisfy bYBE. In this regard, an interesting question is whether there is an associated Yangian symmetry that would constrain the coefficient $k_0$ and the bound-state reflection matrices without the use of the boundary Yang-Baxter equation. Such an ``achiral twisted Yangian'', of a similar structure to the Yangian of the $Y=0$ $D$3-brane introduced in \cite{AN} and explored in \cite{MR2,Palla}, 
will be considered in a forthcoming paper. 

\paragraph{Acknowledgments.} We thank Juan Mart\'{\i}n Maldacena for helpful discussions and for the suggestion to reconsider the integrability of the $D$5-brane boundary conditions. 
We also want to thank Niall MacKay for many valuable discussions and for reading the manuscript and Alessandro Torrielli for lots of useful discussions and comments. V.R. and C.A.S.Y. are supported by UK EPSRC grant EP/H000054/1.

\vfill


\newpage
\appendix

\section{Fundamental $S$-matrix}

The fundamental $S$-matrix may be neatly defined as a differential operator
\begin{equation}
\mathcal{S}(p_{1},p_{2})=a_{i}(p_{1},p_{2})\,\Lambda_{i},
\end{equation}
acting on the superspace with the basis \{$\omega_1$, $\omega_2$, $\theta_3$,
$\theta_4$\}, where $\omega_{a}$ and $\theta_{\alpha}$ are bosonic and fermionic
variables respectively (see \cite{Arutyunov1} for details), and $\Lambda_{i}$ are the
$\mathfrak{su}(2)\oplus\mathfrak{su}(2)$ invariant differential operators:
\begin{align}
\Lambda_{1} &= \frac{1}{2}\left(\omega_{a}^{1}\omega_{b}^{2}+\omega_{b}^{1}\omega_{a}^{2}\right)\frac{\partial^{2}}{\partial\omega_{b}^{2}\partial\omega_{a}^{1}},
&& \Lambda_{6} = \omega_{a}^{2}\theta_{\alpha}^{1}\frac{\partial^{2}}{\partial\omega_{a}^{2}\partial\theta_{\alpha}^{1}}.  \nonumber \\
\Lambda_{2} &= \frac{1}{2}\left(\omega_{a}^{1}\omega_{b}^{2}-\omega_{b}^{1}\omega_{a}^{2}\right)\frac{\partial^{2}}{\partial\omega_{b}^{2}\partial\omega_{a}^{1}},
&& \Lambda_{7}=\epsilon^{ab}\omega_{a}^{1}\omega_{b}^{2}\epsilon_{\alpha\beta}\frac{\partial^{2}}{\partial\theta_{\beta}^{2}\partial\theta_{\alpha}^{1}}, \nonumber \\
\Lambda_{3} &= \frac{1}{2}\left(\theta_{\alpha}^{1}\theta_{\beta}^{2}+\theta_{\beta}^{1}\theta_{\alpha}^{2}\right)\frac{\partial^{2}}{\partial\theta_{\beta}^{2}\partial\theta_{\alpha}^{1}},
&& \Lambda_{8} = \frac{1}{2}\epsilon^{\alpha\beta}\theta_{\alpha}^{1}\theta_{\beta}^{2}\epsilon_{ab}\frac{\partial^{2}}{\partial\omega_{b}^{2}\partial\omega_{a}^{1}}, \nonumber \\
\Lambda_{4} &= \frac{1}{2}\left(\theta_{\alpha}^{1}\theta_{\beta}^{2}-\theta_{\beta}^{1}\theta_{\alpha}^{2}\right)\frac{\partial^{2}}{\partial\theta_{\beta}^{2}\partial\theta_{\alpha}^{1}},
&& \Lambda_{9} = \omega_{a}^{1}\theta_{\alpha}^{2}\frac{\partial^{2}}{\partial\omega_{a}^{2}\partial\theta_{\alpha}^{1}}, \nonumber\\
\Lambda_{5} &= \omega_{a}^{1}\theta_{\alpha}^{2}\frac{\partial^{2}}{\partial\omega_{a}^{1}\partial\theta_{\alpha}^{2}},
&&\Lambda_{10} = \omega_{a}^{2}\theta_{\alpha}^{1}\frac{\partial^{2}}{\partial\omega_{a}^{1}\partial\theta_{\alpha}^{2}},
\label{S_diff_op}
\end{align}
The physical $S$-matrix ($\mathcal{S} : \mathcal{V}_1\otimes\mathcal{V}_2 \mapsto \mathcal{V}_2\otimes\mathcal{V}_1$),
that we were using in our calculations is acquired by acting
with a graded permutation on the $S$-matrix defined on superspace,
$\mathcal{S}_{\text{physical}} := P_{12}\,\mathcal{S}_{\text{superspace}}$.
The coefficients of the fundamental (physical) left $S$-matrix
that we were using in our calculations are:

\begin{align}
a_{1} & =-\frac{x_{1}^{-}-x_{2}^{+}}{x_{2}^{-}-x_{1}^{+}}\frac{\eta_{1}\eta_{2}}{\tilde{\eta}_{1}\tilde{\eta}_{2}},
 && a_{6} =-\frac{x_{1}^{-}-x_{2}^{-}}{x_{2}^{-}-x_{1}^{+}}\frac{\eta_{2}}{\tilde{\eta}_{2}}, \nonumber \\
a_{2} & =-\frac{x_{2}^{+}-x_{1}^{-}}{x_{2}^{-}-x_{1}^{+}}\Biggl( 1-2\frac{1-1/x_{2}^{-} x_{1}^{+}}{1-1/x_{2}^{+} x_{1}^{+}} \frac{x_{2}^{-}-x_{1}^{-}}{x_{2}^{+}-x_{1}^{-}} \Biggr) \frac{\eta_{1}\eta_{2}}{\tilde{\eta}_{1}\tilde{\eta}_{2}},
 && a_{7} =\frac{i\zeta(x_{1}^{-}-x_{1}^{+})(x_{2}^{-}-x_{2}^{+})(x_{1}^{+}-x_{2}^{+})}{(-1+x_{1}^{-}x_{2}^{-})(x_{2}^{-}-x_{1}^{+}) \tilde{\eta}_{1}\tilde{\eta}_{2}}, \nonumber \\
a_{3} & =-1,
 && a_{8} =\frac{i(x_{1}^{-}-x_{2}^{-}) \eta_{1}\eta_{2}}{\zeta(x_{2}^{-}-x_{1}^{+})(-1+x_{1}^{+}x_{2}^{+})}, \nonumber \\
a_{4} & =\Biggl(1-2\frac{1-1/x_{2}^{+} x_{1}^{-}}{1-1/x_{2}^{-} x_{1}^{-}}\frac{x_{2}^{+}-x_{1}^{+}}{x_{2}^{-}-x_{1}^{+}}\Biggr),
 && a_{9} =\frac{x_{1}^{-}-x_{1}^{+}}{-x_{2}^{-}+x_{1}^{+}}\frac{\eta_{2}}{\tilde{\eta}_{1}}, \nonumber \\
a_{5} & =-\frac{x_{1}^{+}-x_{2}^{+}}{x_{2}^{-}-x_{1}^{+}}\frac{\eta_{1}}{\tilde{\eta}_{1}},
 && a_{10} =-\frac{x_{2}^{-}-x_{2}^{+}}{x_{2}^{-}-x_{1}^{+}}\frac{\eta_{1}}{\tilde{\eta}_{2}}.
\end{align}

\section{Reflection matrix $\mathcal{K}^{h}$}
The $K$-matrix for the reflection in the horizontal vacuum case  may be defined on the superspace as
\begin{equation}
\mathcal{K}^{h}(p,-p)=k_{i}(p)\,\Lambda_{i},
\end{equation}
where $\Lambda_{i}$ are the same as in (\ref{S_diff_op}). The reflection coefficients are:
\begin{equation}
k_{i}(p) = a_{i}(p,-p)\,.
\end{equation}

\section{Reflection matrices $\mathcal{K}^{Ba}$ and $\overline{\mathcal{K}}^{Ba}$}

The supersymmetric reflection $K$-matrix $\mathcal{K}^{Ba}$ describing the reflection
of the two-magnon bound states in the bulk from the fundamental states on the boundary
may be defined as a differential operator
\begin{equation}
\mathcal{K}^{Ba}(p_{1},x_B)=k^{\text{\tiny (S)}}_{i}(p_{1},x_B)\,\Lambda_{i}
\end{equation}
acting on the superspace, where $\Lambda_{i}$ are
\begin{align}
\Lambda_{1} &= \frac{1}{6}\left(\omega_{a}^{1}\omega_{b}^{1}\omega_{c}^{2}+\omega_{c}^{1}\omega_{b}^{1}\omega_{a}^{2}+\omega_{a}^{1}\omega_{c}^{1}\omega_{b}^{2}\right)\frac{\partial^{3}}{\partial\omega_{c}^{2}\partial\omega_{b}^{1}\partial\omega_{a}^{1}},
 && \Lambda_{10}=\frac{1}{2}\epsilon^{kl}\omega_{c}^{1}\omega_{k}^{1}\omega_{l}^{2}\epsilon_{\alpha\beta}\frac{\partial^{3}}{\partial\omega_{c}^{2}\partial\theta_{\beta}^{1}\partial\theta_{\alpha}^{1}}, \nonumber \\
\Lambda_{2} &= \frac{1}{6}\left(\epsilon_{bc}\omega_{a}^{1}+\epsilon_{ac}\omega_{b}^{1}\right)\epsilon^{kl}\omega_{k}^{1}\omega_{l}^{2}\frac{\partial^{3}}{\partial\omega_{c}^{2}\partial\omega_{b}^{1}\partial\omega_{a}^{1}},
 && \Lambda_{11}=\frac{1}{2}\epsilon^{\alpha\beta}\omega_{b}^{2}\theta_{\alpha}^{1}\theta_{\beta}^{1}\epsilon_{ac}\frac{\partial^{3}}{\partial\omega_{c}^{2}\partial\omega_{b}^{1}\partial\omega_{a}^{1}}, \nonumber \\
\Lambda_{3} &= \frac{1}{2}\theta_{\beta}^{1}\left(\omega_{a}^{1}\omega_{c}^{2}+\omega_{c}^{1}\omega_{a}^{2}\right)\frac{\partial^{3}}{\partial\omega_{c}^{2}\partial\omega_{a}^{1}\partial\theta_{\beta}^{1}},
 && \Lambda_{12}=\epsilon^{\alpha\beta}\omega_{c}^{2}\theta_{\alpha}^{1}\theta_{\beta}^{1}\epsilon_{\gamma\delta}\frac{\partial^{3}}{\partial\omega_{a}^{1}\partial\theta_{\delta}^{2}\partial\theta_{\gamma}^{1}}, \nonumber \\
\Lambda_{4} &= \frac{1}{2}\theta_{\beta}^{1}\left(\omega_{a}^{1}\omega_{c}^{2}-\omega_{c}^{1}\omega_{a}^{2}\right)\frac{\partial^{3}}{\partial\omega_{c}^{2}\partial\omega_{a}^{1}\partial\theta_{\beta}^{1}},
 && \Lambda_{13}=\epsilon^{\alpha\beta}\omega_{b}^{1}\theta_{\alpha}^{1}\theta_{\beta}^{2}\epsilon_{ac}\frac{\partial^{3}}{\partial\omega_{c}^{2}\partial\omega_{b}^{1}\partial\omega_{a}^{1}}, \nonumber \\
\Lambda_{5} &= \frac{1}{2}\omega_{a}^{1}\omega_{b}^{1}\theta_{\gamma}^{2}\frac{\partial^{3}}{\partial\omega_{b}^{1}\partial\omega_{a}^{1}\partial\theta_{\gamma}^{2}},
 && \Lambda_{14}=\frac{1}{2}\epsilon^{\alpha\beta}\omega_{a}^{2}\theta_{\alpha}^{1}\theta_{\beta}^{1}\epsilon_{\gamma\delta}\frac{\partial^{3}}{\partial\omega_{a}^{1}\partial\theta_{\beta}^{2}\partial\theta_{\gamma}^{1}}, \nonumber \\
\Lambda_{6} &= \frac{1}{2}\omega_{c}^{2}\theta_{\alpha}^{1}\theta_{\beta}^{1}\frac{\partial^{3}}{\partial\omega_{c}^{2}\partial\theta_{\beta}^{1}\partial\theta_{\alpha}^{1}},
 && \Lambda_{15}=\frac{1}{2}\epsilon^{\alpha\beta}\omega_{c}^{1}\theta_{\alpha}^{1}\theta_{\beta}^{2}\epsilon_{\gamma\delta}\frac{\partial^{3}}{\partial\omega_{c}^{2}\partial\theta_{\delta}^{1}\partial\theta_{\gamma}^{1}}, \nonumber \\
\Lambda_{7} &= \frac{1}{2}\omega_{a}^{1}\left(\theta_{\beta}^{1}\theta_{\gamma}^{2}+\theta_{\gamma}^{1}\theta_{\beta}^{2}\right)\frac{\partial^{3}}{\partial\omega_{a}^{1}\partial\theta_{\gamma}^{2}\partial\theta_{\beta}^{1}},
 && \Lambda_{16}=\epsilon^{\alpha\gamma}\theta_{\alpha}^{1}\theta_{\beta}^{1}\theta_{\gamma}^{2}\epsilon_{ac}\frac{\partial^{3}}{\partial\omega_{c}^{2}\partial\omega_{a}^{1}\partial\theta_{\beta}^{1}}, \nonumber \\
\Lambda_{8} &= \frac{1}{2}\omega_{a}^{1}\left(\theta_{\beta}^{1}\theta_{\gamma}^{2}-\theta_{\gamma}^{1}\theta_{\beta}^{2}\right)\frac{\partial^{3}}{\partial\omega_{a}^{1}\partial\theta_{\gamma}^{2}\partial\theta_{\beta}^{1}},
&& \Lambda_{17}=\epsilon^{kl}\omega_{k}^{1}\omega_{l}^{2}\theta_{\beta}^{1}\epsilon_{\alpha\gamma}\frac{\partial^{3}}{\partial\theta_{\gamma}^{2}\partial\theta_{\beta}^{1}\partial\theta_{\alpha}^{1}}, \nonumber \\
\Lambda_{9} &= \theta_{\alpha}^{1}\theta_{\beta}^{1}\theta_{\gamma}^{2}\frac{\partial^{3}}{\partial\theta_{\gamma}^{2}\partial\theta_{\beta}^{1}\partial\theta_{\alpha}^{1}},
 && \Lambda_{18}=\omega_{b}^{1}\omega_{a}^{2}\theta_{\gamma}^{1}\frac{\partial^{3}}{\partial\omega_{b}^{1}\partial\omega_{a}^{1}\partial\theta_{\gamma}^{2}}, \nonumber \\
& && \Lambda_{19}=\omega_{a}^{1}\omega_{c}^{1}\theta_{\beta}^{2}\frac{\partial^{3}}{\partial\omega_{c}^{2}\partial\omega_{a}^{1}\partial\theta_{\beta}^{1}}.\label{LKBa}
\end{align}
The coefficients of the symmetric reflection matrix $\mathcal{K}^{Ba}$ are:
\begin{align}
k_{1}^{\text{\tiny (S)}} & =1\nonumber\\
k_{2}^{\text{\tiny (S)}} & =\frac{3 x_B (x^{-})^2-x_B (x^{+})^2 (2+3 (x^{+})^2)+x^{-}x^{+} (x_B-4 x^{+}+x_B (x^{+})^2)}{2 (x_B+(-1+x_B^2)x^{-}-x_B (x^{-})^2) (x^{+})^2}\nonumber\\
k_{3}^{\text{\tiny (S)}} & =-\frac{((x^{-})^2+x_B x^{+})}{(x_B-x^{-})x^{-}} \frac{\tilde{\eta}}{\eta},\nonumber\\
k_{4}^{\text{\tiny (S)}} & =-\frac{(x_B+x^{+}) (x^{-}+x_B (x^{+})^2)}{(x_B+(-1+x_B^2)x^{-}-x_B (x^{-})^2) x^{+}} \frac{\tilde{\eta}}{\eta},\nonumber\\
k_{5}^{\text{\tiny (S)}} & =\frac{(x_B x^{-}-(x^{+})^2)}{(x_B-x^{-})x^{-}} \frac{\tilde{\eta}_B}{\eta_B},\nonumber\\
k_{6}^{\text{\tiny (S)}} & =\frac{(-x_B (x^{-})^4+x_B (x^{+})^2+x^{-} x^{+}(x_B+x_B (x^{-})^2+x^{-} (4-2 x_B x^{+})))}{2(x^{-})^2 (x_B+(-1+x_B^2) x^{-}-x_B (x^{-})^2)} \frac{\tilde{\eta}^2}{\eta^2},\nonumber\\
k_{7}^{\text{\tiny (S)}} & =-\frac{x^{+} (x_B+x^{+})}{(x_B-x^{-}) x^{-}} \frac{\tilde{\eta} \tilde{\eta}_B}{\eta \eta_B},\nonumber\\
k_{8}^{\text{\tiny (S)}} & =\frac{(2 x_B^2 (x^{-})^3+x_B (x^{-})^2 (x_B-x^{+})x^{+}+2 (x^{+})^3+x^{-} x^{+} (-x_B+x^{+}))}{(x^{-})^2 (x_B+(-1+x_B^2)x^{-}-x_B (x^{-})^2)} \frac{\tilde{\eta}\tilde{\eta}_B}{\eta \eta_B},\nonumber\\
k_{9}^{\text{\tiny (S)}} & =\frac{x^{+} (x_B+x^{+}) (-x_B (x^{-})^2+x^{+})}{(x^{-})^2 (x_B+(-1+x_B^2)x^{-}-x_B (x^{-})^2)} \frac{\tilde{\eta}^2 \tilde{\eta}_B}{\eta^2 \eta_B},\nonumber\\
k_{10}^{\text{\tiny (S)}} & =\frac{i \zeta ((x^{-})^2-(x^{+})^2)^2 (x_B x^{+}+x_B(x^{-})^2 x^{+}+x^{-} (x_B+2 x^{+}-x_B (x^{+})^2))}{4(x^{-})^2 (x_B+(-1+x_B^2) x^{-}-x_B (x^{-})^2) x^{+} (-1+x_B x^{+})} \frac{1}{\eta^2},\nonumber\\
k_{11}^{\text{\tiny (S)}} & =-\frac{i x_B (x^{-}+x^{+})^2}{2 \zeta (x_B-x^{-})x^{-} (1+x_B x^{-}) x^{+}} \tilde{\eta}^2,\nonumber\\
k_{12}^{\text{\tiny (S)}} & =-\frac{i \zeta x_B (x_B x^{-}-(x^{+})^2) ((x^{-})^2-(x^{+})^2)}{\sqrt{2}x^{-} (x_B+(-1+x_B^2) x^{-}-x_B (x^{-})^2) x^{+}} \frac{1}{\eta \eta_B},\nonumber\\
k_{13}^{\text{\tiny (S)}} & =\frac{i (x^{-}+x^{+}) (x_B x^{-}-(x^{+})^2)}{\sqrt{2} \zeta (x_B-x^{-})x^{-} (1+x_B x^{-}) x^{+}} \tilde{\eta} \tilde{\eta}_B, \nonumber\\
k_{14}^{\text{\tiny (S)}} & =\frac{x_B (x_B (x^{-})^2-x^{+}) (x^{-}+x^{+})}{\sqrt{2} (x^{-})^2 (-x_B+x^{-}) (1+x_B x^{-})} \frac{\tilde{\eta}^2}{\eta \eta_B}, \nonumber\\
k_{15}^{\text{\tiny (S)}} & =\frac{(x_B (x^{-})^2-x^{+}) ((x^{-})^2-(x^{+})^2)}{\sqrt{2} (x^{-})^2(x_B+(-1+x_B^2) x^{-}-x_B (x^{-})^2)} \frac{\tilde{\eta} \tilde{\eta}_B}{\eta^2}, \nonumber\\
k_{16}^{\text{\tiny (S)}} & =-\frac{i (x_B+x^{+}) (x^{-}+x^{+})}{\sqrt{2} \zeta x^{-} (-x_B+x^{-})(1+x_B x^{-})} \frac{\tilde{\eta}^2\tilde{\eta}_B}{\eta},\nonumber\\
k_{17}^{\text{\tiny (S)}} & =\frac{i \zeta x_B (x_B+x^{+}) (-(x^{-})^2+(x^{+})^2)}{\sqrt{2} x^{-} (x_B+(-1+x_B^2) x^{-}-x_B (x^{-})^2)} \frac{\tilde{\eta}}{\eta^2 \eta_B},\nonumber\\
k_{18}^{\text{\tiny (S)}} & =\frac{x_B (x^{-}+x^{+}) }{\sqrt{2} (x_B-x^{-})x^{-}} \frac{\tilde{\eta}}{\eta_B}, \nonumber\\
k_{19}^{\text{\tiny (S)}} & =\frac{((x^{-})^2-(x^{+})^2)}{\sqrt{2}x^{-} (-x_B+x^{-})} \frac{\tilde{\eta}_B}{\eta}.\label{ks}
\end{align}

The anti-supersymmetric reflection $K$-matrix $\overline{\mathcal{K}}^{Ba}$ describing the reflection
of the two-magnon bound states in the mirror bulk theory from the fundamental states on the boundary
may be defined as a differential operator
\begin{equation}
\overline{\mathcal{K}}^{Ba}(p_{1},x_B)=k^{\text{\tiny (A)}}_{i}(p_{1},x_B)\,\overline{\Lambda}_{i}
\end{equation}
acting on the mirror superspace, where $\overline{\Lambda}_{i}$ are the differential operators
acting on the mirror superspace. They may be acquired from (\ref{LKBa}) by interchange of
bosonic and fermionic indices, $(a,\;b) \leftrightarrow (\alpha,\;\beta)$.
The reflection coefficients $k^{\text{\tiny (A)}}_{i}$ may be obtained from \ref{ks} using the relation
$k^{\text{\tiny (A)}}_{i}(p,x_B)=k^{\text{\tiny (S)}}_{i}(-p,x_B)$.

\section{1-loop computations in the dCFT}
\label{1lopp}

The 1-loop mixing matrix of anomalous dimensions for the scalar sector of the defect conformal field theory (dCFT) was obtained in \cite{DeMa}. Let us recall that the complex scalar defect fields $\phi^a$ transform in a $\mathbf{2}$ of ${\mathfrak so}(3)_H$, while the six scalar fields of the bulk theory are split into $X_H^I$ and $X_V^A$, transforming in a $\mathbf{3}$ of ${\mathfrak so}(3)_H$ and in a $\mathbf{3}$ of ${\mathfrak so}(3)_V$ respectively.

In this appendix we are interested in the case when the reference state breaks the ${\mathfrak so}(3)_V$ symmetry. For definiteness let us take $\bar \phi_a Z\cdots Z \phi^b$, with $Z=X_V^2 + i X_V^3$, as the vacuum state. As already discussed, the 16 bulk impurities will be accommodated into a $\boxslash\hspace{-0.9mm}\boxslash$ and a $\twobv$ of the diagonal $\mathfrak{su}(2|2)_D$. In particular, of the 4 scalar field impurities left after fixing the vacuum, the $X_H^I$ give rise to the $\phi^{\{a,b\}}$ components of the symmetric representation and $X_V^1$ to the $\phi^{[a,b]}$ component of the antisymmetric representation.

In particular, the component $\phi^{\{1,1\}}$ is the combination $Y=X_H^1 + i X_H^2$, and its boundary reflection shall be diagonal if $\phi^1$ defect fields (and their conjugate) are placed at the ends of the chain. Let us consider the superposition of left-moving single magnon $Y$ with a right-moving one.
\begin{equation}
|\Psi^{Y}\rangle = \sum_{n=1}^L (e^{ipn} + { K }^{Y}_{L}(p)e^{-ipn}) |n\rangle, \qquad |n\rangle \equiv
|\bar \phi_2 Z^{n-1}YZ^{L-n}\phi^1\rangle.
\end{equation}
This  superposition  is an eigenstate with eigenvalue $8g^2\sin^2(\tfrac{p}{2})+8g^2$ of the Hamiltonian given in \cite{DeMa} if
\begin{equation}
{ K }^{Y}_{ L}(p) = - \frac{1-3 e^{ip}}{1-3 e^{-ip}}, \qquad  e^{2ip(L+1)} = ({ K }^{Y}_{ L}(p))^2.
\end{equation}

Another impurity that it is reflected diagonally is $X_V^1$. In that case, the superposition
\begin{equation}
|\Psi^{X_V^1}\rangle = \sum_{n=1}^L (e^{ipn} + { K }^{X_V^1}_{{L}}(p)e^{-ipn}) |n\rangle, \qquad |n\rangle \equiv
|\bar \phi_2 Z^{n-1}X_V^1 Z^{L-n}\phi^1\rangle.
\end{equation}
is an eigenstate with eigenvalue $8g^2\sin^2(\tfrac{p}{2})+8g^2$ if
\begin{equation}
{ K }^{X_V^1}_{{L}}(p) =  e^{ip}, \qquad  e^{2ip(L+1)} = ({ K }^{X_V^1}_{{L}}(p))^2.
\end{equation}


\end{document}